\documentclass[11pt,fleqn]{article}

\usepackage{a4}
\usepackage{amstext}
\usepackage{amsfonts}
\usepackage{amssymb}
\usepackage{color}
\usepackage{cite}
\usepackage{epsfig}
\usepackage{mathtools}
\usepackage{ulem}
\usepackage{caption}         
\usepackage{subcaption}   
\usepackage{framed}
\usepackage{multirow}
\usepackage{ulem}

\usepackage{authblk}

\newcommand{\Eqref}[1]{Eq.~\eqref{#1}}

\newcommand{\gtapprox}{\raisebox{-0.5ex}{$\,\stackrel{>}{\scriptstyle\sim}\,$}}
\newcommand{\ltapprox}{\raisebox{-0.5ex}{$\,\stackrel{<}{\scriptstyle\sim}\,$}}

\begin{document}


\begin{center}

{\huge \bf Determination of $\Lambda_{\overline{\textrm{MS}}}^{(n_f=2)}$ and analytic}

{\huge \bf parameterization of the static}

\vspace{0.2cm}
{\huge \bf quark-antiquark potential}

\vspace{0.5cm}

\textbf{Felix Karbstein$^{1,2}$, Marc Wagner$^3$ and Michelle Weber$^3$}

$^1$~Helmholtz-Institut Jena, Fr\"obelstieg 3, D-07743 Jena, Germany

$^2$~Theoretisch-Physikalisches Institut, Friedrich-Schiller-Universit\"at Jena, Max-Wien-Platz 1, D-07743 Jena, Germany

$^3$~Goethe-Universit\"at Frankfurt am Main, Institut f\"ur Theoretische Physik, Max-von-Laue-Stra{\ss}e 1, D-60438 Frankfurt am Main, Germany

\vspace{0.5cm}

May 13, 2018

\end{center}

\begin{tabular*}{16cm}{l@{\extracolsep{\fill}}r} \hline \end{tabular*}

\vspace{-0.4cm}
\begin{center} \textbf{Abstract} \end{center}
\vspace{-0.4cm}

While lattice QCD allows for reliable results at small momentum transfers (large quark separations), perturbative QCD is restricted to large momentum transfers (small quark separations). The latter is determined up to a reference momentum scale $\Lambda$, which is to be provided from outside, e.g.\ from experiment or lattice QCD simulations. In this article, we extract $\Lambda_{\overline{\textrm{MS}}}$ for QCD with $n_f=2$ dynamical quark flavors by matching the perturbative static quark-antiquark potential in momentum space to lattice results in the intermediate momentum regime, where both approaches are expected to be applicable. In a second step, we combine the lattice and the perturbative results to provide a complete analytic parameterization of the static quark-antiquark potential in position space up to the string breaking scale. As an exemplary phenomenological application of our all-distances potential we compute the bottomonium spectrum in the static limit.

\begin{tabular*}{16cm}{l@{\extracolsep{\fill}}r} \hline \end{tabular*}

\thispagestyle{empty}


\newpage

\setcounter{page}{1}

\section{\label{sec:intro} Introduction}

This article finalizes our attempts started in \cite{Jansen:2011vv}, and subsequently improved in \cite{Karbstein:2014bsa}, to determine $\Lambda_{\overline{\textrm{MS}}}$ by matching the static quark-antiquark potential obtained form perturbation theory to lattice data.
While this endeavor seems almost trivial on first sight, it actually is not and requires to deal with several problems and tricky issues. 
To keep our series of articles self-contained and to allow for a fair comparison of the different approaches and their results, also in the present article we stick to quantum chromodynamics (QCD) with $n_f=2$ dynamical quark flavors. However, the described approach can be readily adopted to other setups. 

Note that many other studies, resorting to different lattice QCD ensembles generated with different numbers of dynamical quark flavors, have pursued related strategies to extract $\Lambda_{\overline{\textrm{MS}}}$ from the static quark-antiquark potential \cite{Michael:1992nj,Gockeler:2005rv,Brambilla:2010pp,Leder:2010kz,Bazavov:2012ka,Fritzsch:2012wq,Bazavov:2014soa}. 
Besides, various complementary approaches exist to determine $\Lambda_{\overline{\rm MS}}$, or alternatively the strong coupling $\alpha_s$ at a specific momentum scale.
For recent results based on lattice computations, see, e.g., \cite{DellaMorte:2004bc,Shintani:2008ga,Aoki:2009tf,Sternbeck:2010xu,McNeile:2010ji,Blossier:2010ky,
Sternbeck:2012qs,Blossier:2013ioa,Cichy:2013eoa,
Brida:2016flw,DallaBrida:2016kgh,Bruno:2017gxd}, employing the Schr\"odinger functional, vacuum polarization functions, ghost and gluon propagators, heavy quark correlators and the Dirac operator spectrum. Other works study $\tau$ decays, the collision of electrons with positrons and protons or holographic QCD \cite{Abbate:2010xh,Kneur:2011vi,Boito:2012cr,Gehrmann:2012sc,Abbas:2012fi,Kneur:2013coa,Alekhin:2013nda,
Andreev:2014wwa,Deur:2014qfa,Deur:2016opc}. For a recent review concerning the QCD running coupling cf.\ \cite{Deur:2016tte}.

Perturbation theory is based on series expansions in the strong coupling $\alpha_s$, which thus is required to be small.
The physical coupling $\alpha_s\equiv\alpha_s(\mu)$ generically depends on a momentum scale $\mu$, which can be considered as a measure of the typical momentum transfer in a given process. Due to asymptotic freedom of QCD, $\alpha_s(\mu)\ll 1$ for large values of $\mu$, while $\alpha_s(\mu)\gg1$ for small values of $\mu$.
In momentum space we have $\mu\sim p$, while in position space $\mu \sim 1/r$.
Correspondingly, perturbative calculations of the static potential in QCD are limited to small quark-antiquark separations $r$ or large relative momentum transfers $p$, respectively.
They are conventionally carried out in momentum space, where the static potential is presently known up to ${\cal O}(\alpha_s^4)$.
Lattice simulations are tailored to the manifestly non-perturbative regime of QCD.
They are naturally performed in position space and allow for controlled insights into the static potential from a minimum distance of a few times the lattice spacing $a$.
Hence, the basic idea to determine $\Lambda_{\overline{\textrm{MS}}}$ from the quark-antiquark potential amounts to fitting the perturbative static potential to lattice data in the intermediate regime of separations (momentum transfers) where both approaches are expected to allow for trustworthy insights, using $\Lambda_{\overline{\textrm{MS}}}$ as fitting parameter.

In our initial article \cite{Jansen:2011vv}, we pursued this strategy in position space.
To this end we transformed the perturbative static potential from momentum space to position space via an ordinary Fourier transform.
As this Fourier transform also receives contributions from momenta where the perturbative expression is no longer trustworthy, the resulting perturbative potential in position space suffers from uncontrolled contributions, significantly worsening its convergence behavior as compared to the original momentum space potential.  
It can, however, be shown that the introduction of an additional momentum scale can remove these pathologies \cite{Aglietti:1995tg,Jezabek:1998pj,Jezabek:1998wk,Beneke:1998rk,Hoang:1998nz,Beneke:1998ui,Pineda:2001zq,Pineda:2002se}, thereby facilitating a reliable extraction of $\Lambda_{\overline{\textrm{MS}}}$.
Proceeding along these lines, in \cite{Jansen:2011vv} we obtained $\Lambda_{\overline{\textrm{MS}}}^{(n_f=2)}=315(30)\,{\rm MeV}$.

Subsequently, in \cite{Karbstein:2014bsa} we argued that this analysis can be performed more reliably in momentum space.
To obtain the lattice potential in momentum space, we employed a discrete Fourier transform in three dimensions.
In order to increase the number of available data points for the discrete Fourier transform, governing the resolution in momentum space, we used a fitting function to extrapolate the long distance behavior inferred from the lattice simulations up to distances of several hundreds of the lattice spacing $a$.
The values of the extrapolating function were then stored for the sites of our extended three dimensional lattice serving as supplementary input for the discrete Fourier transform.
Pursuing this strategy, in \cite{Karbstein:2014bsa} we found $\Lambda_{\overline{\textrm{MS}}}^{(n_f=2)}=331(21)\,{\rm MeV}$.
Note, that the results of both analyses \cite{Jansen:2011vv} and \cite{Karbstein:2014bsa} are compatible with each other, the latter exhibiting a smaller error.

In the present article, we further improve and streamline the procedure to extract $\Lambda_{\overline{\textrm{MS}}}$ in momentum space.
The main difference to \cite{Karbstein:2014bsa} is that we do not perform a discrete Fourier transform of lattice data from position to momentum space.
Instead, we immediately parameterize the discrete lattice data points for the static potential in position space by a continuous function, thereby providing us with a ``continuous lattice potential''.
More specifically, we choose the parameterizing function such that its Fourier transform to momentum space can be performed analytically, implying that the construction of the lattice potential in momentum space becomes essentially trivial.
Another improvement to our previous articles \cite{Jansen:2011vv} and \cite{Karbstein:2014bsa} is that in the meantime the knowledge of the QCD $\beta$-function has been improved by an additional order in $\alpha_s$ \cite{Baikov:2016tgj}.
This results in a more precise relation between the strong coupling $\alpha_s(\mu)$ and the dimensionless ratio $\mu/\Lambda_{\overline{\textrm{MS}}}$, prospectively further diminishing the error of our extracted value for $\Lambda_{\overline{\textrm{MS}}}$.

However, the present article does not only aim at an accurate and efficient determination of $\Lambda_{\overline{\textrm{MS}}}$.
The extracted value of $\Lambda_{\overline{\textrm{MS}}}$ is immediately used to construct a complete analytic parameterization of the static quark-antiquark potential in position space up to the string breaking scale.
This potential encodes both perturbative and manifestly non-perturbative information and has various phenomenological applications.
For instance, it can be employed to study the spectrum of heavy quarkonia. As an example, here we adopt it to bottomonium in the static limit.

More specifically, our article is organized as follows.

Sec.~\ref{SEC068} is devoted to the lattice computation of the static quark-antiquark potential $V_\text{lat}$ for QCD with $n_f=2$ dynamical quark flavors in position space.
Here, our main goal is to parameterize the discrete data points for the potential obtained from Wilson loop averages in Sec.~\ref{sec:Wilsonloop} by a continuous function.
In Sec.~\ref{SEC462}, we confirm that a simple three parameter fit of the Cornell form, $V_\text{lat} = V_0 - \alpha/r + \sigma r$, with offset $V_0$ and parameters $\alpha$ and $\sigma$, already accurately describes the lattice data points.
We do not only extract the values of $\alpha$ and $\sigma$ and their errors, but also account for their correlations. 

Sec.~\ref{SEC067} focuses on the perturbative static potential.
After briefly reviewing the known contributions to the static potential in momentum space $\tilde{V}_\text{pert}$ in Sec.~\ref{sec:Vpertmom}, we discuss its position space analogue $V_\text{pert}$
in Sec.~\ref{sec:Vpertpos}.
Sec.~\ref{sec:alpha} details on the relation of the perturbative strong coupling $\alpha_s(\mu)$ at a given momentum scale $\mu$ to $\Lambda_{\overline{\text{MS}}}$.
In general, the constituting equation relating the dimensionless ratio $\mu/\Lambda_{\overline{\text{MS}}}$ to $\alpha_s(\mu)$ cannot be solved analytically for $\alpha_s(\mu)$.
However, an analytical expression for $\alpha_s(\mu)$ can be extracted in the limit of $\mu/\Lambda_{\overline{\text{MS}}}\gg1$.

In Sec.~\ref{SEC547} we determine $\Lambda_{\overline{\text{MS}}}$ for QCD with $n_f=2$ dynamical quark flavors.
To this end, we fit $\tilde{V}_\text{pert}$ to the Fourier transform of the continuous lattice potential $V_\text{lat}$ in the intermediate momentum regime where both lattice simulations and perturbation theory are expected to allow for trustworthy results, treating $\Lambda_{\overline{\text{MS}}}$ as fitting parameter.

Using the extracted value of $\Lambda_{\overline{\text{MS}}}$, in Sec.~\ref{SEC421} we construct a complete analytic parameterization of the static potential in position space $V$, interpolating between both the perturbative and the manifestly non-perturbative regime.
To this end, $V_\text{pert}$ is smoothly connected to the continuous lattice potential $V_\text{lat}$ at intermediate quark-antiquark separations.
In order to allow for an analytic representation of $V_\text{pert}$, we employ the analytic expression for $\alpha_s(\mu)$ in the limit of $\mu/\Lambda_{\overline{\text{MS}}}\gg 1$.

In Sec.~\ref{SEC566} we study the bottomonium spectrum in the static limit. 
This analysis serves as an application of the all-distances potential $V$ constructed in the preceding section.

Finally, we end with conclusions and a brief outlook in Sec.~\ref{sec:concl}.


\newpage

\section{\label{SEC068}Lattice computation of the static potential}


\subsection{\label{SEC008}Lattice setup}

We use the same $n_f = 2$ gauge link configurations as in our previous articles concerned with the determination of $\Lambda_{\overline{\textrm{MS}}}^{(n_f=2)}$ \cite{Jansen:2011vv,Karbstein:2014bsa}. These gauge link configurations were generated by the European Twisted Mass Collaboration (ETMC) \cite{Boucaud:2007uk,Boucaud:2008xu,Baron:2009wt}. The gluon action is the tree-level Symanzik improved gauge action \cite{Weisz:1982zw},
\begin{eqnarray}
S_\textrm{gluon}[U]  =  \frac{\beta}{6} \bigg(b_0 \sum_{x,\mu\neq\nu} \textrm{Tr}\Big(1 - P^{1 \times 1}(x;\mu,\nu)\Big) + b_1 \sum_{x,\mu\neq\nu} \textrm{Tr}\Big(1 - P^{1 \times 2}(x;\mu,\nu)\Big)\bigg)
\end{eqnarray}
with $b_0 = 1 - 8 b_1$ and $b_1 = -1/12$, and the quark action is the Wilson twisted mass action \cite{Frezzotti:2000nk,Frezzotti:2003ni,Frezzotti:2004wz,Shindler:2007vp},
\begin{eqnarray}
\label{EQN963} S_\textrm{quark}[\chi,\bar{\chi},U]  =  \sum_x \bar{\chi}(x) \Big(D_{\rm W} + i \mu_\mathrm{q} \gamma_5 \tau_3\Big) \chi(x)
\end{eqnarray}
with
\begin{eqnarray}
D_\mathrm{W}  =  \frac{1}{2} \Big(\gamma_\mu \Big(\nabla_\mu + \nabla^\ast_\mu\Big) - \nabla^\ast_\mu \nabla_\mu\Big) + m_0 .
\end{eqnarray}
$\nabla_\mu$ and $\nabla^\ast_\mu$ are the gauge covariant forward and backward derivatives, $m_0$ and $\mu_\mathrm{q}$ are the bare untwisted and twisted quark masses, $\tau_3$ is the third Pauli matrix acting in flavor space, and $\chi = (\chi^{(u)} , \chi^{(d)})$ represents the quark fields in the so-called twisted basis.

The twist angle $\omega$ is given by $\omega = \arctan(\mu_\mathrm{R} / m_\mathrm{R})$, where $\mu_\mathrm{R}$ and $m_\mathrm{R}$ denote the renormalized twisted and untwisted quark masses. For the ensembles of gauge link configurations considered in the present study $\omega$ has been tuned to $\pi / 2$ by adjusting $m_0$ appropriately. This ensures automatic $\mathcal{O}(a)$ improvement for many observables including the static potential (cf.\ \cite{Boucaud:2008xu} for details).

The considered gauge link configurations cover several different values of the lattice spacing (cf.~Table~\ref{TAB077}, which also provides the corresponding pion masses $m_\textrm{PS}$, spacetime volumes $(L/a)^3 \times T/a$ and numbers of gauge link configurations used for the computations of the static potential). The lattice spacing in physical units has been set via the pion mass and the pion decay constant using chiral perturbation theory. The resulting value for the hadronic scale\footnote{The hadronic scale $r_0$ is defined via $r_0^2 F(r_0) = 1.65$, with $F(r) = {\rm d}V(r) / {\rm d}r$ \cite{Sommer:1993ce}.} $r_0$ is $r_0 = 0.420(14) \, \textrm{fm}$ (cf.\ also Sec.~5 of \cite{Boucaud:2008xu} and Table~8 of \cite{Baron:2009wt}). For further details on the generation of these gauge field configurations as well as on the computation and the analysis of standard quantities, such as lattice spacing and pion mass, we refer to \cite{Boucaud:2008xu,Baron:2009wt}.




 










\begin{table}[htb]
\begin{center}
\begin{tabular}{|c|c|c|c|c|c|}
\hline
 & & & & & \vspace{-0.40cm} \\
$\beta$ & $a$ in $\textrm{fm}$ & $(L/a)^3 \times T/a$ & $m_\textrm{PS}$ in $\textrm{MeV}$ & $r_0 / a$ & \# gauges no-HYP/HYP \\
 & & & & & \vspace{-0.40cm} \\
\hline
 & & & & & \vspace{-0.40cm} \\
\hline
 & & & & & \vspace{-0.40cm} \\
$3.90$ & $0.079(3)\phantom{00}$ & $24^3 \times 48$ & $340(13)$ & $5.36(4)\phantom{0}$ & $168/108$ \\
 & & & & & \vspace{-0.40cm} \\
\hline
 & & & & & \vspace{-0.40cm} \\
$4.05$ & $0.063(2)\phantom{00}$ & $32^3 \times 64$ & $325(10)$ & $6.73(5)\phantom{0}$ & $\phantom{0}71/189$ \\
 & & & & & \vspace{-0.40cm} \\
\hline
 & & & & & \vspace{-0.40cm} \\
$4.20$ & $0.0514(8)\phantom{0}$ & $24^3 \times 48$ & $284(5)\phantom{0}$ & $8.36(6)\phantom{0}$ & $123/211$ \\
 & & & & & \vspace{-0.40cm} \\
\hline
 & & & & & \vspace{-0.40cm} \\
$4.35$ & $0.0420(17)$           & $32^3 \times 64$ & $352(22)$ & $9.81(13)$ & $146/295$\vspace{-0.40cm} \\
 & & & & & \\
\hline
\end{tabular}
\caption{\label{TAB077}Ensembles of gauge link configurations.}
\end{center}
\end{table}


\subsection{\label{sec:Wilsonloop}Extracting the lattice static potential from Wilson loop averages}

We extract the static potential in position space $V_\textrm{lat}(\vec{r})$ from the exponential decay of Wilson loop averages $\langle W(\vec{r},t) \rangle$ with respect to their temporal extent $t$, while keeping their spatial extent $\vec{r}$ fixed \cite{Brown:1979ya}. To this end we first compute
\begin{eqnarray}
V_\textrm{lat}^\textrm{(effective)}(\vec{r},t)  =  \frac{1}{a} \ln\bigg(\frac{\langle W(\vec{r},t) \rangle}{\langle W(\vec{r},t+a) \rangle}\bigg) .
\end{eqnarray}
In a second step the $t$-independent quantity $V_\textrm{lat}(\vec{r})$ is obtained by performing an uncorrelated $\chi^2$ minimizing fit to $V_\textrm{lat}^\textrm{(effective)}(\vec{r},t)$ in a suitable $t$ range. This range is chosen such that the contributions of excited states are strongly suppressed, while statistical errors are still small.

We perform two independent computations on each of the ensembles listed in Table~\ref{TAB077}.
\begin{itemize}
\item \textit{no-HYP} computation: \\
Temporal links remain unchanged, i.e.\ are not smeared. The resulting static potential has small discretization errors, in particular at small quark-antiquark separations $r=|\vec{r}|$, but large statistical errors at large separations. To obtain a fine resolution at small $r$, we consider both on- and off-axis Wilson loops (for a detailed explanation regarding the construction of off-axis Wilson loops, cf.\ \cite{Jansen:2011vv}). Spatial links are APE smeared, to improve the ground state overlap and, hence, to be able to extract the static potential more precisely ($N_\textrm{APE} = 20$, $\alpha_\textrm{APE} = 0.5$; cf. \cite{Jansen:2008si} for details). Besides, the tree-level improvement technique put forward by \cite{Necco:2001xg,Necco:2003jf} is employed to further reduce discretization errors.

\item \textit{HYP} computation: \\
Temporal links are HYP2 smeared, which corresponds to using the HYP2 static quark action \cite{Hasenfratz:2001hp,DellaMorte:2003mn,DellaMorte:2005nwx}. The resulting static potential has large discretization errors at small quark-antiquark separations, but the reduced self energy of the static quarks leads to significantly smaller statistical errors at large separations. We consider only on-axis Wilson loops. Spatial links are again APE smeared ($N_\textrm{APE} = 60$, $\alpha_\textrm{APE} = 0.5$).
\end{itemize}
While the \textit{no-HYP} results have already been used in our previous determinations of $\Lambda_{\overline{\textrm{MS}}}^{(n_f=2)}$ \cite{Jansen:2011vv,Karbstein:2014bsa}, the \textit{HYP} results have been generated for this study.


\subsection{\label{SEC462}Parameterization of the discrete lattice data by a continuous function}

In contrast to our previous determination of $\Lambda_{\overline{\textrm{MS}}}^{(n_f=2)}$ in momentum space \cite{Karbstein:2014bsa}, we parameterize the discrete lattice data for the static potential by a continuous function before transforming to momentum space. This has several advantages. For example rotational symmetry is restored already at an early stage, thereby avoiding technical problems like performing a cylinder cut.
Moreover, this approach is technically less complicated and the uncertainty of the final result for $\Lambda_{\overline{\textrm{MS}}}$ is somewhat smaller; see below.
Finally, the continuous function used to parameterize the lattice potential forms an important constituent of the analytic all-distances potential constructed in Sec.~\ref{SEC421}.

\begin{figure}[htb]
\begin{center}

\begin{picture}(0,0)%
\includegraphics{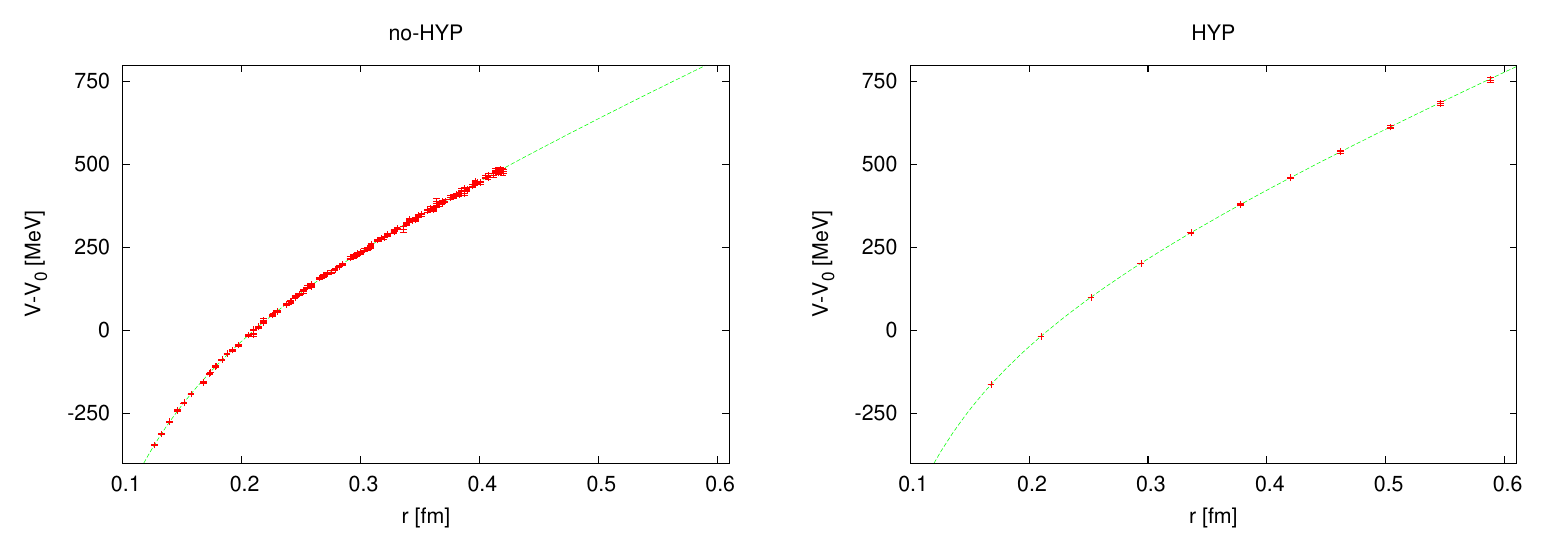}%
\end{picture}%
\setlength{\unitlength}{4144sp}%
\begingroup\makeatletter\ifx\SetFigFont\undefined%
\gdef\SetFigFont#1#2#3#4#5{%
  \reset@font\fontsize{#1}{#2pt}%
  \fontfamily{#3}\fontseries{#4}\fontshape{#5}%
  \selectfont}%
\fi\endgroup%
\begin{picture}(7110,2456)(1,-1617)
\end{picture}%

\caption{\label{FIG008}Results of \textit{no-HYP} (left) and \textit{HYP} (right) computations of the lattice static potential at $\beta=4.35$ (red data points).
The dashed green curves show the corresponding analytic parameterizations~\eqref{EQN170}, with parameters $\alpha = \overline{\alpha} = 0.347$, $\sigma = \overline{\sigma} = 7.86 / \textrm{fm}^2$ ($\alpha = \overline{\alpha} = 0.353$, $\sigma = \overline{\sigma} = 7.55 / \textrm{fm}^2$) in the left (right) plot; cf. Table~\ref{TAB766}.}
\end{center}
\end{figure}

For quark-antiquark separations $r_\textrm{min} \leq r \leq r_\textrm{max}$, with $r_\textrm{min} \gtapprox 0.13 \, \textrm{fm}$ and $r_\textrm{max} \leq 0.79 \, \textrm{fm}$ (the maximum range of separations, where results are available) the lattice potential computed on all four ensembles can be parameterized consistently by the Cornell potential,
\begin{eqnarray}
\label{EQN170} V_\textrm{lat}(r)  =  V_0 - \alpha \frac{1}{r} + \sigma r .
\end{eqnarray}
Here, $V_0$ is a constant shift of the potential and $\sigma$ is the string tension.
Even though $V_0$ amounts to a physically irrelevant shift within lattice QCD alone, it will be explicitly needed in Sec.~\ref{SEC421} below to match the static potentials obtained from lattice QCD and perturbation theory.
While $\alpha = \pi/12 \approx 0.26$ for large $r$ in the bosonic string picture \cite{Luscher:1980fr,Luscher:1980ac}, lattice simulations with $n_f=2$ quark flavors have extracted $\alpha \approx +0.3 \ldots +0.5$ \cite{Donnellan:2010mx}, which is in agreement with our results. We have explicitly checked, whether accounting for additional terms $\sim\ln^m(r) / r$ with $m\in\{1,2,3\}$ and $\sim1/r^m$ with $m\in\{2,3\}$ improves the parameterization of the static potential~\eqref{EQN170}. We find that at our current level of statistical precision such terms are not needed. Their prefactors are zero within statistical errors.
Moreover, none of these terms has the potential to reduce $\chi^2_\textrm{red}$ significantly with respect to the values quoted in Table~\ref{TAB766}, implying that their inclusion does not enhance the parameterization of the static potential. This finding is also supported by Figure~\ref{FIG008}, depicting \textit{no-HYP} and \textit{HYP} results for our smallest lattice spacing together with the corresponding analytic parameterizations~\eqref{EQN170}. The latter accurately describe the lattice data points and no systematic deviations are visible.

\begin{table}[htb]
\begin{center}
\begin{tabular}{|c|c||c|c|c|c||c|c|c|c|c|}
\hline
 & & & & & & & & & & \vspace{-0.40cm} \\
 & & & & \# data & & & & & & \\
 & $X$ & $r_\textrm{min} / a$ & $r_\textrm{max} / a$ & points & $\chi_\textrm{red}^2$ & $\overline{\alpha}$ & $\Delta \alpha$ & $\overline{\sigma} \, \textrm{fm}^2$ & $\Delta \sigma \, \textrm{fm}^2$ & $\textrm{corr}(\alpha,\sigma)$ \\
 & & & & & & & & & & \vspace{-0.40cm} \\
\hline
 & & & & & & & & & & \vspace{-0.40cm} \\
$\beta=3.90$ & \textit{no-HYP} & $2.83$ & $8.00$ & $\phantom{0}72$ & $0.36$ & $0.414$ & $0.006$ & $7.94$ & $0.09$ & $-0.88$ \\
               &    \textit{HYP} & $3$    & $10$ & $\phantom{00}8$ & $0.34$ & $0.415$ & $0.015$ & $7.31$ & $0.19$ & $-0.96$ \\
 & & & & & & & & & & \vspace{-0.40cm} \\
$\beta=4.05$ & \textit{no-HYP} & $2.83$ & $8.00$ & $\phantom{0}72$ & $0.53$ & $0.386$ & $0.006$ & $7.87$ & $0.08$ & $-0.89$ \\
               &    \textit{HYP} & $3$    & $10$ & $\phantom{00}8$ & $1.78$ & $0.391$ & $0.008$ & $7.39$ & $0.15$ & $-0.96$ \\
 & & & & & & & & & & \vspace{-0.40cm} \\
$\beta=4.20$ & \textit{no-HYP} & $2.83$ & $10.0$ & $126$ & $0.66$ & $0.368$ & $0.006$ & $7.60$ & $0.10$ & $-0.76$ \\
               &    \textit{HYP} & $3$    & $12$ & $\phantom{0}10$ & $0.12$ & $0.382$ & $0.011$ & $7.41$ & $0.18$ & $-0.92$ \\
 & & & & & & & & & & \vspace{-0.40cm} \\
$\beta=4.35$ & \textit{no-HYP} & $3.00$ & $10.0$ & $124$ & $0.70$ & $0.347$ & $0.004$ & $7.86$ & $0.09$ & $-0.85$ \\
               &    \textit{HYP} & $4$    & $14$ & $\phantom{0}11$ & $0.22$ & $0.353$ & $0.007$ & $7.55$ & $0.15$ & $-0.93$ \\
 & & & & & & & & & & \vspace{-0.40cm} \\
\hline
 & & & & & & & & & & \vspace{-0.40cm} \\
continuum & \textit{no-HYP} & & & & & $0.324$ & $0.006$ & $7.47$ & $0.56$ & $-0.19$ \\
&    \textit{HYP} & & & & & $0.330$ & $0.011$ & $7.57$ & $0.57$ & $-0.25$ \\
 & & & & & & & & & & \vspace{-0.40cm} \\
\hline
 & & & & & & & & & & \vspace{-0.40cm} \\
continuum & combined & & & & & $0.326$ & $0.005$ & $7.52$ & $0.55$ & $-0.17$\vspace{-0.40cm} \\
 & & & & & & & & & & \\
\hline
\end{tabular}
\caption{\label{TAB766}Parameterization of the lattice static potential via \Eqref{EQN170}. To allow for a straightforward comparison, the results for $\sigma$ at fixed $\beta$ (upper eight rows) have been converted from lattice units to $1 / \textrm{fm}^2$ without accounting for the lattice spacing errors. The column ``\# data points'' gives the number of data points available for the extraction of $V_\text{lat}(r)$.}
\end{center}
\end{table}

The parametrization (\ref{EQN170}) does not account for string breaking, which is happening at quark-antiquark separations $r \approx r_\textrm{sb}$, where $V(r_\textrm{sb}) = 2 m_B$, with $m_B$ denoting the mass of the lightest heavy-light  meson (quantum numbers $J^P = 0^- , 1^-$; cf.\ e.g.\ \cite{Jansen:2008si,Michael:2010aa}). The string breaking distance has been determined using lattice QCD in \cite{Bali:2005fu}, yielding $r_{\textrm{sb}} = 1.13(10)(10) \, \textrm{fm}$. In this section we do not use any lattice data for $r > r_\textrm{sb}$, and hence do not consider string breaking; cf. Sec.~\ref{SEC566} for a study of the effect of string breaking on the bottomonium spectrum.

In the present study, we determine the parameters $V_0$, $\alpha$ and $\sigma$ by performing uncorrelated $\chi^2$ minimizing fits of \Eqref{EQN170} to the discrete lattice QCD data of the static potential for separations $r_\textrm{min} \leq r \leq r_\textrm{max}$; cf. Table~\ref{TAB766} for the explicit values of $r_{\rm min}$ and $r_{\rm max}$. The minimum distance $r_\textrm{min}$ is needed to exclude data points with sizable lattice discretization errors, and $r_\textrm{max}$ is required to exclude unwanted artifacts of the spatial periodicity of the lattice. In detail we proceed as follows:
\begin{itemize}
\item[(i)] For each of the four ensembles characterized by $\beta$ and both $X \in \{ \textit{no-HYP} , \textit{HYP} \}$ computations, we determine the averages $\overline{\alpha}^{\beta,X}$ and $\overline{\hat{\sigma}}^{\beta,X}$, where $\hat{\sigma} = \sigma a^2$ is the string tension in units of the lattice spacing.
The errors $\Delta \alpha^{\beta,X}$ and $\Delta \hat{\sigma}^{\beta,X}$ are computed via the jackknife method.
Moreover, we determine the correlation $\textrm{corr}(\alpha^{\beta,X},\hat{\sigma}^{\beta,X})$. For two generic quantities $\alpha$ and $\sigma$, the latter is defined as
\begin{eqnarray}
\textrm{corr}(\alpha,\sigma)  =  \frac{\langle (\alpha - \overline{\alpha}) (\sigma - \overline{\sigma}) \rangle}{\sqrt{\langle (\alpha - \overline{\alpha})^2 \rangle \langle (\sigma - \overline{\sigma})^2 \rangle}}\,.
\end{eqnarray}
The respective results are collected in Table~\ref{TAB766}, together with the corresponding values of $\chi^2_\textrm{red} = \chi^2 / \textrm{dof}$. Here, dof counts the degrees of freedom, $\textrm{dof} = (\#\ \textrm{data points}) - 3$, with ``$\#\ \textrm{data points}$'' denoting the number of data points available for the extraction of $V_\text{lat}(r)$ and $3$ representing the number of fit parameters. Note that results obtained for the same $\beta$ but different $X \in \{ \textit{no-HYP} , \textit{HYP} \}$ may disagree within statistical errors, because of different discretization errors.

\item[(ii)] For both \textit{no-HYP} and \textit{HYP} results we perform continuum extrapolations of $\alpha$ and $\sigma$ to linear order in $a^2$, which is the leading order of discretization errors in Wilson twisted mass lattice QCD at maximal twist (cf.\ Sec.~\ref{SEC008}). Correlations of $\alpha$ and $\sigma$ are properly taken into account by using jackknife samples $(\alpha , \sigma)$ from step~(i). Moreover, we account for the lattice spacing errors listed in Table~\ref{TAB077} when converting a dimensionless $\hat{\sigma}$ to a dimensionful $\sigma$. Since the lattice spacing errors constitute the dominant source of uncertainty for $\sigma$, they also reduce the correlation $\textrm{corr}(\alpha,\sigma)$ significantly; cf.\ the upper eight rows to the lower three rows of Table~\ref{TAB766}. Furthermore, notice that pion masses and spacetime volumes for different lattice spacings are similar, but not identical (cf.\ Table~\ref{TAB077}).
We do not consider this as problematic, since the dependence of the potential on the pion mass and the spacetime volume is negligible within statistical errors \cite{Jansen:2011vv}. This is also supported by the small reduced $\chi^2$ of the continuum extrapolations of both our \textit{no-HYP} and \textit{HYP} computations. These extrapolations are shown in Figure~\ref{FIG002}, and the corresponding results for $\overline{\alpha}^X$, $\Delta \alpha^X$, $\overline{\sigma}^X$, $\Delta \sigma^X$ and $\textrm{corr}(\alpha^X,\sigma^X)$ are collected in the 9th and 10th row of Table~\ref{TAB766}. As expected, the continuum extrapolated \textit{no-HYP} and \textit{HYP} results for both $\alpha$ and $\sigma$ are in agreement within statistical errors.

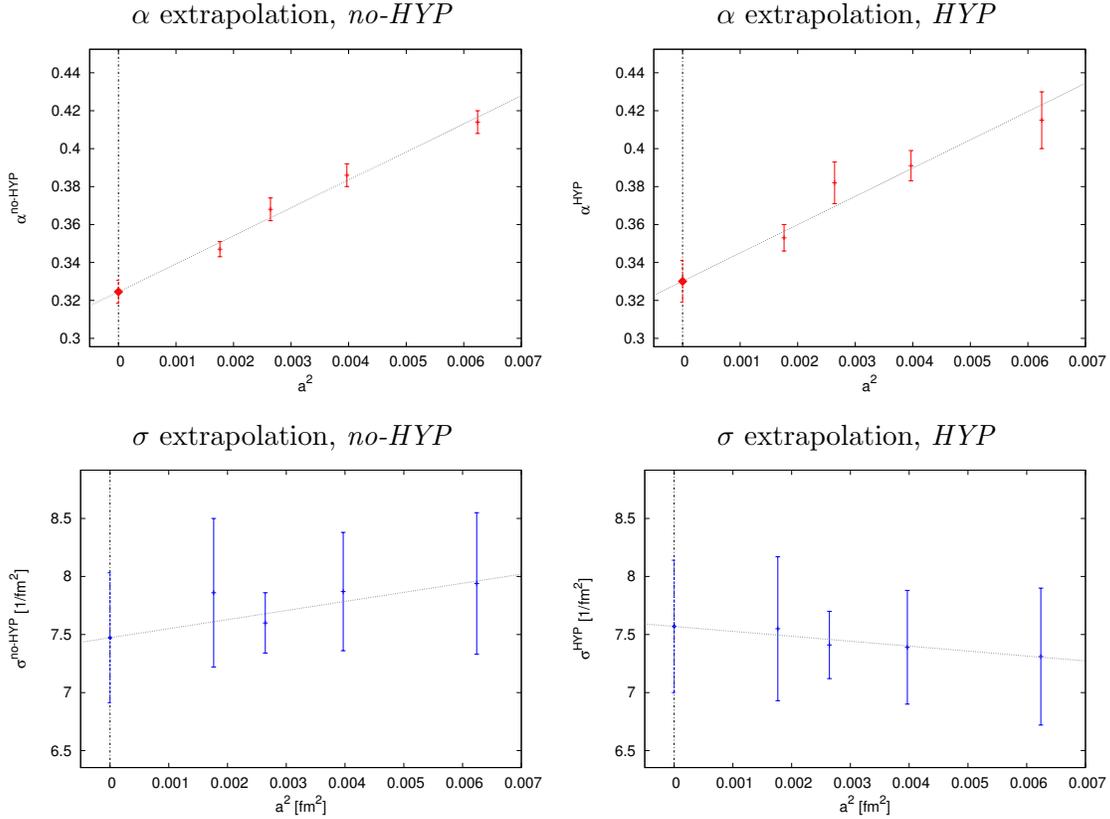
\begin{figure}[htb]
\begin{center}

\begin{picture}(0,0)%
\includegraphics{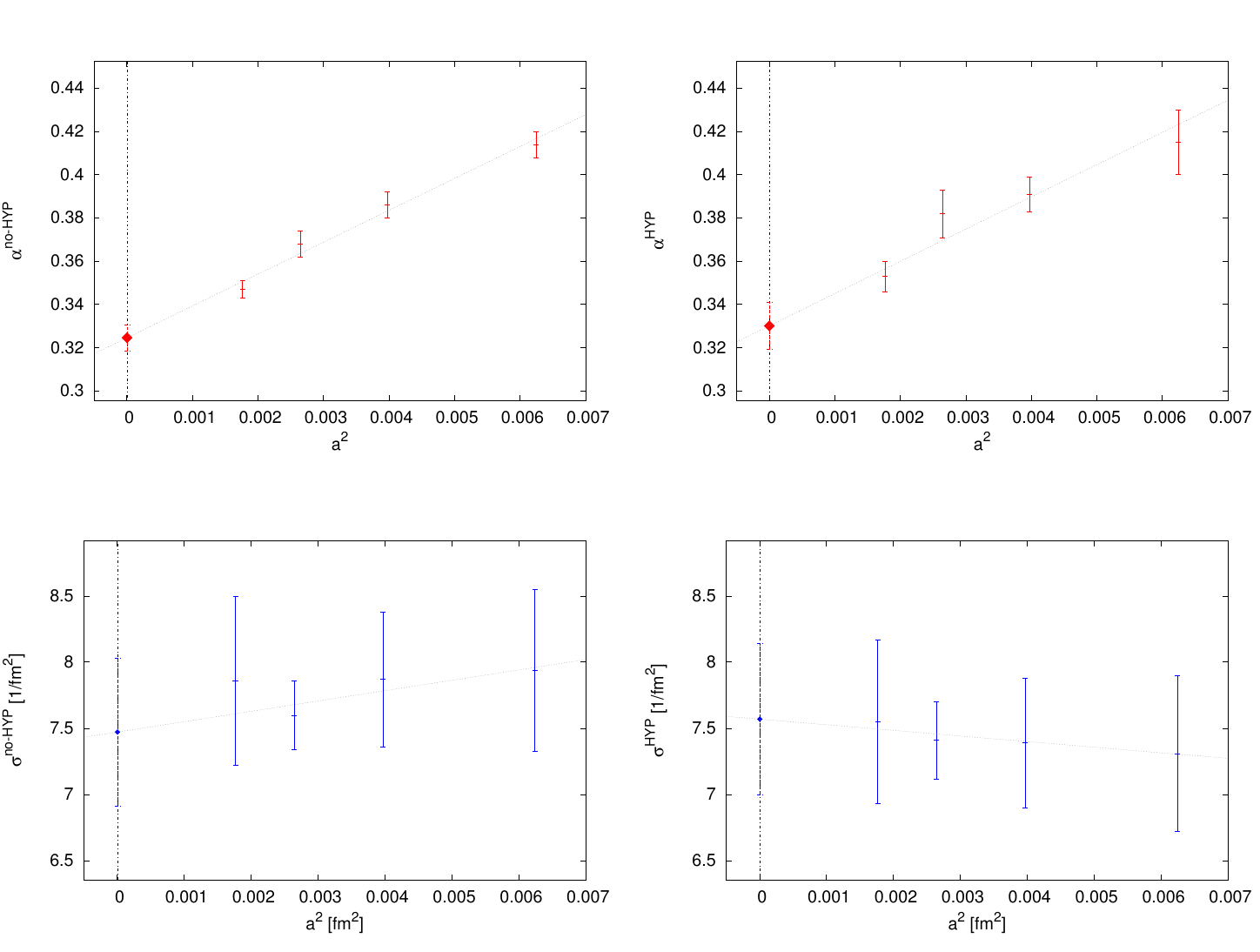}%
\end{picture}%
\setlength{\unitlength}{4144sp}%
\begingroup\makeatletter\ifx\SetFigFont\undefined%
\gdef\SetFigFont#1#2#3#4#5{%
  \reset@font\fontsize{#1}{#2pt}%
  \fontfamily{#3}\fontseries{#4}\fontshape{#5}%
  \selectfont}%
\fi\endgroup%
\begin{picture}(6615,4991)(1,-3948)
\put(1711,884){\makebox(0,0)[b]{\smash{{\SetFigFont{11}{13.2}{\familydefault}{\mddefault}{\updefault}{\color[rgb]{0,0,0}$\alpha$ extrapolation, \textit{no-HYP}}%
}}}}
\put(5086,884){\makebox(0,0)[b]{\smash{{\SetFigFont{11}{13.2}{\familydefault}{\mddefault}{\updefault}{\color[rgb]{0,0,0}$\alpha$ extrapolation, \textit{HYP}}%
}}}}
\put(5086,-1636){\makebox(0,0)[b]{\smash{{\SetFigFont{11}{13.2}{\familydefault}{\mddefault}{\updefault}{\color[rgb]{0,0,0}$\sigma$ extrapolation, \textit{HYP}}%
}}}}
\put(1711,-1636){\makebox(0,0)[b]{\smash{{\SetFigFont{11}{13.2}{\familydefault}{\mddefault}{\updefault}{\color[rgb]{0,0,0}$\sigma$ extrapolation, \textit{no-HYP}}%
}}}}
\end{picture}%

\caption{\label{FIG002}Continuum extrapolation of the parameters $\alpha$ and $\sigma$ in the analytic parameterization~\eqref{EQN170} of the lattice static potential for both \textit{no-HYP} and \textit{HYP} computations.}
\end{center}
\end{figure}

\item[(iii)] We combine the continuum extrapolated \textit{no-HYP} and \textit{HYP} results from step~(ii) by performing constant fits. Correlations between $\alpha$ and $\sigma$ are properly taken into account by using jackknife samples $(\alpha , \sigma)$ from step~(ii). The fits are shown in Figure~\ref{FIG003}, and the results for $\overline{\alpha}$, $\Delta \alpha$, $\overline{\sigma}$, $\Delta \sigma$ and $\textrm{corr}(\alpha,\sigma)$ are collected in Table~\ref{TAB766}.

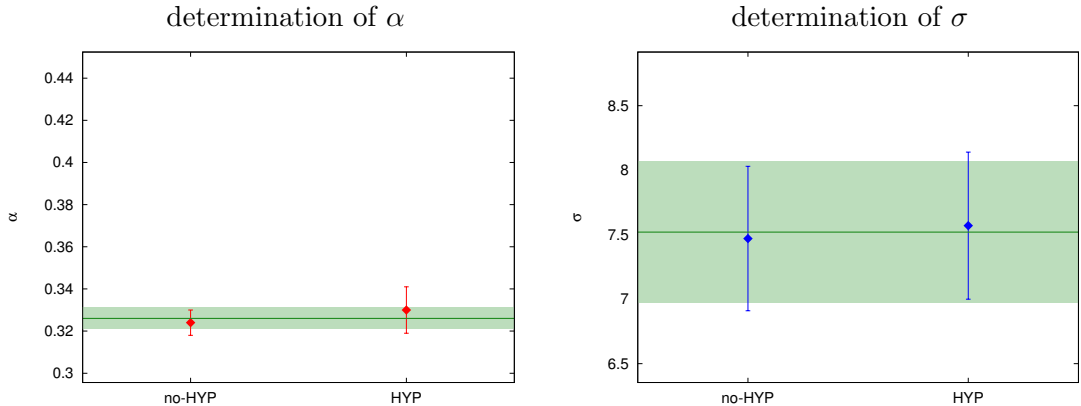
\begin{figure}[htb]
\begin{center}

\begin{picture}(0,0)%
\includegraphics{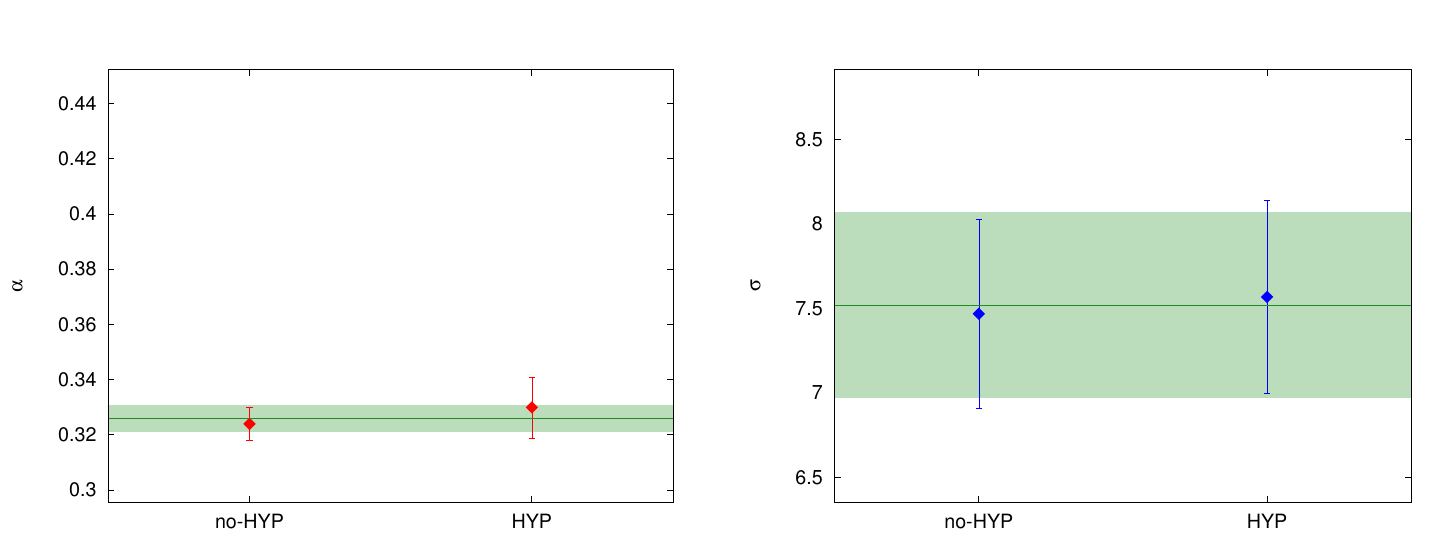}%
\end{picture}%
\setlength{\unitlength}{4144sp}%
\begingroup\makeatletter\ifx\SetFigFont\undefined%
\gdef\SetFigFont#1#2#3#4#5{%
  \reset@font\fontsize{#1}{#2pt}%
  \fontfamily{#3}\fontseries{#4}\fontshape{#5}%
  \selectfont}%
\fi\endgroup%
\begin{picture}(6615,2471)(1,-1428)
\put(1711,884){\makebox(0,0)[b]{\smash{{\SetFigFont{11}{13.2}{\familydefault}{\mddefault}{\updefault}{\color[rgb]{0,0,0}determination of $\alpha$}%
}}}}
\put(5086,884){\makebox(0,0)[b]{\smash{{\SetFigFont{11}{13.2}{\familydefault}{\mddefault}{\updefault}{\color[rgb]{0,0,0}determination of $\sigma$}%
}}}}
\end{picture}%

\caption{\label{FIG003}Combination of continuum extrapolated \textit{no-HYP} and \textit{HYP} results. This determines our final values of the parameters $\alpha$ and $\sigma$ provided in the last row of Table~\ref{TAB766}.
The dark green lines correspond to our final values for $\overline{\alpha}$ and $\overline{\sigma}$, and the green bands depict our final errors $\Delta\alpha$ and $\Delta\sigma$, respectively.}
\end{center}
\end{figure}
\end{itemize}

The values for ``$\Delta \alpha$'' and ``$\Delta \sigma \, \textrm{fm}^2$'' in Table~\ref{TAB766} imply that the \textit{no-HYP} results constrain both $\alpha$ and $\sigma$ more precisely than the \textit{HYP} results. At first glance this might seem somewhat surprising, since lattice QCD results for the static potential computed with HYP smeared temporal links are generically much more precise than analogous results obtained without HYP smearing. The important point to note here is that the \textit{no-HYP} computations comprise not only on-axis Wilson loops but also all possible off-axis Wilson loops, while the \textit{HYP} computations are exclusively based on on-axis Wilson loops (cf.\ Sec.~\ref{sec:Wilsonloop}). In turn, much more data points for $V_\text{lat}(r)$ are available for \textit{no-HYP} computations (cf.\ the column ``\# data points'' in Table~\ref{TAB766}), resulting in smaller errors for $\alpha$ and $\sigma$ in comparison to the \textit{HYP} results. Noteworthily, the \textit{HYP} results nevertheless reduce the final uncertainty of $\alpha$ by roughly 15\% in comparison to the \textit{no-HYP} results alone.
On the other hand, the uncertainties of the continuum extrapolations of the \textit{no-HYP}, \textit{HYP} and combined results for $\sigma$ differ only marginally, because they are strongly dominated by the lattice spacing error (cf.\ \cite{Boucaud:2008xu,Baron:2009wt} for technical details about how this error is determined). Consequently, the precision of $\sigma$ could not be enhanced by including off-axis Wilson loops to the \textit{HYP} data sets.
In any case, we consider it as reassuring to have two independently computed datasets, \textit{no-HYP} and \textit{HYP}, yielding perfectly compatible results.

\begin{figure}[htb]
\begin{center}

\begin{picture}(0,0)%
\includegraphics{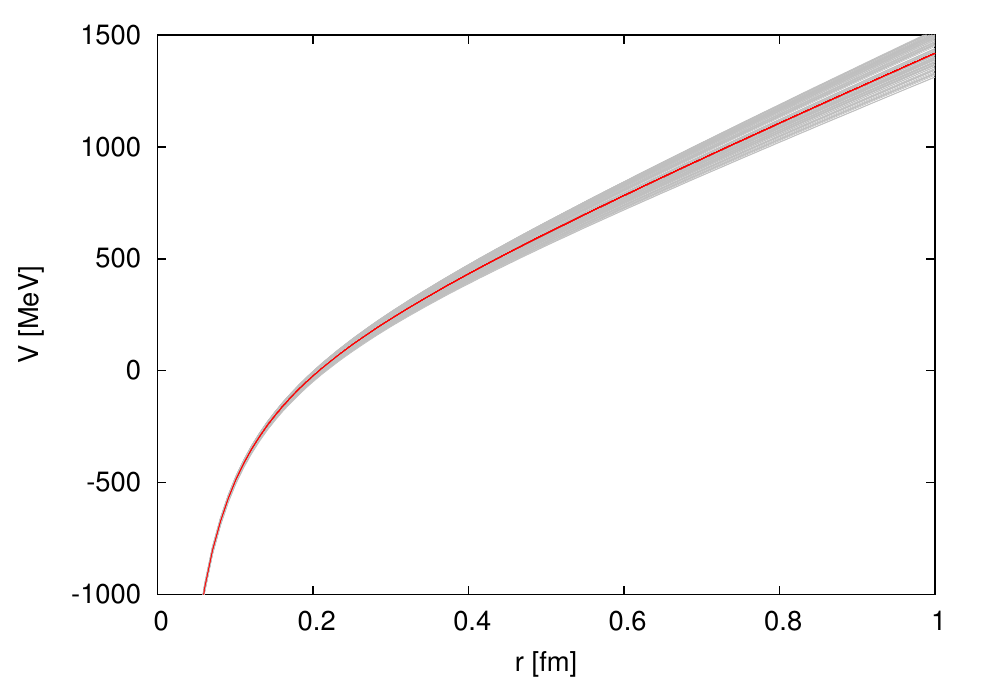}%
\end{picture}%
\setlength{\unitlength}{4144sp}%
\begingroup\makeatletter\ifx\SetFigFont\undefined%
\gdef\SetFigFont#1#2#3#4#5{%
  \reset@font\fontsize{#1}{#2pt}%
  \fontfamily{#3}\fontseries{#4}\fontshape{#5}%
  \selectfont}%
\fi\endgroup%
\begin{picture}(4500,3149)(1,-2310)
\end{picture}%

\caption{\label{FIG004}Analytic parameterization~\eqref{EQN170} of the lattice potential.
The red curve corresponds to $V_0 = 0$, $\alpha = \overline{\alpha} = 0.325$ and $\sigma = \overline{\sigma} = 7.51 / \textrm{fm}^2$ (cf.\ last row of Table~\ref{TAB766}), while the gray error band has been generated from the jackknife samples of step~(iii).}
\end{center}
\end{figure}

The continuum extrapolation of the analytic parameterization~\eqref{EQN170} of the lattice static potential combining both {\it no-HYP} and {\it HYP} computations (cf.\ last row of Table~\ref{TAB766}) is shown in Figure~\ref{FIG004}.

To cross-check our results with previous ETMC analyses, we determine the continuum extrapolated $r_0$ from our results for $\alpha$ and $\sigma$ via
\begin{eqnarray}
r_0  =  \bigg(\frac{1.65 - \alpha}{\sigma}\bigg)^{1/2} ,
\end{eqnarray} 
yielding $r_0 = 0.420(15)$. This value is in perfect agreement with $r_0 = 0.420(14)$ (extrapolated to the continuum and the chiral limit) from \cite{Baron:2009wt}.

We note that there is an anti-correlation between $\alpha$ and $\sigma$, cf.\ $\textrm{corr}(\alpha,\sigma) = -0.17$ for our final continuum result. This can be explained by the fact that both increasing $\alpha$ and $\sigma$ results in a larger slope of $V_\textrm{lat}(r)$, implying that these two parameters have a similar effect on the shape of $V_\textrm{lat}(r)$. Hence, for precise statistical analyses based on the static potential, e.g.\ the determination of $\Lambda_{\overline{\textrm{MS}}}$ or the computation of the bottomonium spectrum, as done in Sec.~\ref{SEC547} and Sec.~\ref{SEC566}, this anti-correlation should be taken into account. In Sec.~\ref{SEC421} we discuss in detailed how to include this anti-correlation in the computation of any observable, which makes use of the analytic parameterization~\eqref{EQN170} of the static quark-antiquark potential.


\clearpage

\section{\label{SEC067}The perturbative static potential}

\subsection{\label{sec:Vpertmom}The perturbative static potential in momentum space}

In perturbation theory the static quark-antiquark potential $V$ is conventionally determined in momentum space. For gauge group ${\rm SU}(3)$, it can be expressed as 
\begin{equation}
 \tilde{V}_\text{pert}(p)=-\frac{4}{3}\frac{4\pi}{p^2}\alpha_V\bigl(\alpha_s(\mu),\ln(\mu^2/p^2)\bigr) \,, \label{eq:V(p)}
\end{equation}
with $p=|\vec{p}|>\Lambda_{\rm QCD}$. The latter condition implies an explicit restriction to the perturbative momentum regime of QCD. 
The dimensionless quantity $\alpha_V$ is a function of both the strong coupling $\alpha_s(\mu)$ evaluated at the renormalization scale $\mu$ (cf. also Sec.~\ref{sec:alpha} below) and $\ln(\mu^2/p^2)$.
It is explicitly known up to ${\cal O}(\alpha_s^4)$.

The static potential is a renormalization group invariant, implying invariance of $V$ under a change of $\mu$.
In perturbation theory this means that when evaluating a result known up to ${\cal O}(\alpha^{\bar k})$ for two different choices of $\mu$, the differences among these results are relegated to ${\cal O}(\alpha^{\bar k+1})$,
such that for small enough $\alpha_s$ and large enough $\bar k$ the specific choice of $\mu$ eventually becomes irrelevant.

Note, that if $\mu$ is chosen as $\mu=c p$, where $c$ denotes a proportionality constant, the logarithm in the argument of $\alpha_V$ becomes independent of $p$.
For $c=1$ we have $\ln(\mu^2/p^2)|_{\mu=cp}=0$, such that \Eqref{eq:V(p)} can be written in a particularly compact form and $\alpha_V$ becomes a function of $\alpha_s(p)$ only.
Adopting the latter choice, the known terms of the static potential can be represented as 
\begin{equation}
 \tilde{V}_\text{pert}(p)=-\frac{4}{3}\frac{4\pi \alpha_s(p)}{p^2}\biggl[1+\frac{\alpha_s(p)}{4\pi}a_1+\biggl(\frac{\alpha_s(p)}{4\pi}\biggr)^2a_2+\biggl(\frac{\alpha_s(p)}{4\pi}\biggr)^3\Bigl(a_{3{\rm ln}}\ln\alpha_s(p)+a_3\Bigr)\biggr] \,. \label{eq:V(p)_known}
\end{equation}
In the remainder of this article we will refer to \Eqref{eq:V(p)_known}, utilizing the identification $\mu=p$, as the static potential in momentum space.
A truncation of \Eqref{eq:V(p)_known} accounting for terms up to ${\cal O}(\alpha_s^{1+n})$ is subsequently referred to as (next-to-)$^n$leading-order or N$^n$LO, respectively.
The coefficients $a_1$ \cite{Fischler:1977yf,Billoire:1979ih} $a_2$ \cite{Peter:1996ig,Peter:1997me,Schroder:1998vy} and $a_{3{\rm ln}}$ \cite{Appelquist:1977es,Brambilla:1999qa} are known analytically, while
some contributions to $a_3$ are only known numerically \cite{Smirnov:2008pn,Smirnov:2009fh,Anzai:2009tm,Smirnov:2010zc,Anzai:2010td}.
For gauge group ${\rm SU}(3)$, $n_f=2$ dynamical massless quark flavors and in the $\overline{\rm MS}$ scheme \cite{Bardeen:1978yd,Furmanski:1981cw}, they read
\begin{equation}
 a_1=\frac{73}{9}\,, \quad a_2=\frac{25139}{162}+9\pi^2\Bigl(4-\frac{\pi^2}{4}\Bigr)+\frac{94}{3}\zeta(3)\,,\quad a_{3{\rm ln}}=144\pi^2 \,, \quad a_3=8783.16(38)\,. 
\end{equation}

The running of $\alpha_s(\mu)$ with $\mu$ is governed by the QCD $\beta$-function,
\begin{equation}
 \beta\bigl(\alpha_s(\mu)\bigr)=\frac{\mu}{\alpha_s(\mu)}\frac{d \alpha_s(\mu)}{d \mu}\,. \label{eq:beta}
\end{equation}
Its perturbative expansion in $\alpha_s$ is presently known with the following accuracy,
\begin{equation}
 \beta(\alpha_s)=-2\beta_0\sum_{i=0}^4\biggl(\frac{\alpha_s}{4\pi}\biggr)^{1+i}b_i \, , \label{eq:betaseries}
\end{equation}
where $b_i=\beta_i/\beta_0$.
For ${\rm SU}(3)$ with $n_f=2$ and in the $\overline{\rm MS}$ scheme, the coefficients in \Eqref{eq:betaseries} are given by \cite{vanRitbergen:1997va,Baikov:2016tgj}%
\footnote{Note that our conventions slightly differ from those of \cite{Baikov:2016tgj}. In particular, $\beta_i|_\text{\cite{Baikov:2016tgj}}=\beta_i/4^{i+1}$.}
\begin{align}
 \beta_0&=\frac{29}{3}\,, \quad b_1=\frac{230}{29}\,,\quad b_2=\frac{48241}{522}\,,\quad b_3=\frac{18799309}{14094}+\frac{275524}{783}\zeta(3)\,, \nonumber\\
 b_4&=\frac{2522305027}{112752}+\frac{109354687}{4698}\zeta(3)-\frac{68881}{1620}\pi^4-\frac{16675240}{783}\zeta(5)\,. \label{eq:bs}
\end{align}

Equations~\eqref{eq:beta} and \eqref{eq:betaseries} imply that $\alpha_s(p)$ can be expressed in terms of $\alpha_s(\mu)$ and $\beta_0\ln(\mu^2/p^2)$.
Formally expanding $\alpha_s(p)$ in powers of $\alpha_s(\mu)$ and solving these equations order by order in $\alpha_s(\mu)$ one obtains
\begin{align}
 \alpha_s(p)=\alpha_s(\mu)\biggl\{1&+\frac{\alpha_s(\mu)}{4\pi}\beta_0\ln(\mu^2/p^2)\biggl[1+\frac{\alpha_s(\mu)}{4\pi}\Bigl(\beta_0\ln(\mu^2/p^2)+b_1\Bigr) \nonumber\\
 &+\biggl(\frac{\alpha_s(\mu)}{4\pi}\biggr)^2\biggl(\bigl[\beta_0\ln(\mu^2/p^2)\bigr]^2+\frac{5}{2}b_1\beta_0\ln(\mu^2/p^2)+b_2\biggr)
 \biggr]\biggr\}+ {\cal O}(\alpha_s^5)\,. \label{eq:alphas}
\end{align}
With the help of this identity the couplings in \Eqref{eq:V(p)_known} can be promoted to any other renormalization scale.
Upon insertion into \Eqref{eq:V(p)_known}, we recover the structure of \Eqref{eq:V(p)}, with $\alpha_V$ known explicitly up to ${\cal O}(\alpha_s^4)$.

\subsection{\label{sec:Vpertpos}The perturbative static potential in position space}

When choosing $\mu$ as independent of $p$, \Eqref{eq:V(p)} can be straightforwardly transformed to position space by means of a standard Fourier transform in three dimensions,
\begin{equation}
V_\text{pert}(r)=\int\frac{{\rm d}^3p}{(2\pi)^3}\,{\rm e}^{i\vec{p}\cdot\vec{r}}\,\tilde{V}_\text{pert}(p)\,. \label{eq:Fint}
\end{equation}
However, note that the Fourier integral~\eqref{eq:Fint} naturally includes momenta $p\lesssim\Lambda_{\rm QCD}$ for which perturbation theory is no longer trustworthy, potentially inducing uncontrolled contributions.

Contributions of this kind are already present in the perturbative potential $\tilde{V}_\text{pert}(p)$.
In standard perturbation theory loop diagrams come along with integrations $\int\frac{{\rm d}^4q}{(2\pi)^4}$ of the loop four-momentum $q$ over the full momentum regime, i.e.,
also receive contributions from outside the perturbative momentum regime.
The leading uncontrolled contribution to $\tilde{V}_\text{pert}(p)$ is quadratic in $\Lambda_{\rm QCD}$ and scales as $\sim-\frac{4\pi\alpha_s}{p^2}(\frac{\Lambda_{\rm QCD}}{p})^2$ \cite{Beneke:1998rk},
translating to a term $\sim-\frac{\alpha_s}{r}(r\Lambda_{\rm QCD})^2$ in $V_\text{pert}(r)$.
Obviously, the condition for momentum transfers $p$ to be described reliably in perturbation theory, $\frac{\Lambda_{\rm QCD}}{p}\ll 1$, corresponds to the restriction $r\Lambda_{\rm QCD}\ll 1$ in position space.

Contrarily, the leading uncontrolled contribution arising from the Fourier integral~\eqref{eq:Fint} is just linear in $\Lambda_{\rm QCD}$ and scales as $\sim-\frac{\alpha_s}{r}(r\Lambda_{\rm QCD})$ \cite{Beneke:1998rk}, which implies that the Fourier transform to position space in fact enhances the pathological terms, and renders the perturbative expansion of $V_\text{pert}(r)$ worse behaved than that of $\tilde{V}_\text{pert}(p)$ \cite{Aglietti:1995tg,Beneke:1998rk,Hoang:1998nz,Beneke:1998ui,Pineda:2001zq,Pineda:2002se}. In the literature, the contribution $\sim-\frac{\alpha_s}{r}(r\Lambda_{\rm QCD})$ is often referred to as the leading renormalon ambiguity of the perturbative static potential in position space.

However, the latter problem can be cured by manifestly restricting the Fourier integral to the perturbative momentum regime $p\geq\mu_f>\Lambda_{\rm QCD}$, where $\mu_f$ denotes a momentum cutoff still in the perturbative regime; cf. \cite{Beneke:1998rk,Laschka:2011zr,Karbstein:2013zxa}.
The position space potential as defined by a restricted Fourier transform,
\begin{equation}
 V_\text{pert}(r,\mu_f)=\int_{p\geq\mu_f}\frac{{\rm d}^3p}{(2\pi)^3}\,{\rm e}^{i\vec{p}\cdot\vec{r}}\,\tilde{V}_\text{pert}(p)
          =V_\text{pert}(r)- \underbrace{\int_{p<\mu_f}\frac{{\rm d}^3p}{(2\pi)^3}\,{\rm e}^{i\vec{p}\cdot\vec{r}}\,\tilde{V}_\text{pert}(p)}_{=:\delta V_\text{pert}(r,\mu_f)}\,, \label{eq:deltaV} 
\end{equation}
does not suffer from enhanced pathological terms in comparison to $V(p)$.
In this article, we will adopt this definition of the perturbative potential in position space. 
Equation~\eqref{eq:deltaV} can be evaluated analytically \cite{Karbstein:2013zxa}.

To allow for a compact representation of the explicit expressions for $V_\text{pert}(r)$ and $\delta V_\text{pert}(r,\mu_f)$,
it is convenient to introduce the following polynomials of degree $k$ \cite{Chishtie:2001mf},
\begin{equation}
 P_k(L)=\sum_{m=0}^{k}\rho_{km}L^m\,, \label{eq:Pk}
\end{equation}
with dimensionless expansion coefficients $\rho_{km}$.
For $1\leq k\leq3$ we have $\rho_{k0}=a_k$,
\begin{gather}
 \rho_{21}=(2a_1+b_1)\beta_0\,, \quad
 \rho_{31}=(3a_2+2a_1b_1+b_2)\beta_0\,, \quad
 \rho_{32}=(3a_1+\tfrac{5}{2}b_1)\beta_0^2\,,
\end{gather}
and $\rho_{kk}=\beta_0^k$.
Moreover, we use the shorthand notation $P_k^{(n)}(L)=\frac{\partial^n}{\partial L^n} P_k(L)$.

Setting the renormalization scale to $\mu=1/r$, we obtain
\begin{align}
V_\text{pert}(r)=-\frac{4}{3}\,\alpha_s(1/r)\,\frac{1}{r}
&\biggl\{
1+\frac{\alpha_s(1/r)}{4\pi}\Bigl(1+2\gamma_E \beta_0\Bigr) \nonumber\\
&\hspace*{-2mm}+\biggl(\frac{\alpha_s(1/r)}{4\pi}\biggr)^2\biggl[a_2
+2\gamma_E \rho_{21} + \biggl(4\gamma_E^2+\frac{\pi^2}{3}\biggr)\beta_0^2\biggr] \nonumber\\
&\hspace*{-2mm}+\biggl(\frac{\alpha_s(1/r)}{4\pi}\biggr)^3\biggl[a_{3{\rm ln}}\ln\alpha_s+a_3+2\gamma_E \rho_{31}
+\biggl(4\gamma_E^2+\frac{\pi^2}{3}\biggr)\rho_{32} \nonumber\\
&\hspace*{4.9cm}+\Bigl(\bigl(4\gamma_E^2+\pi^2\bigr)\gamma_E+8\zeta(3)\Bigr)2\beta_0^3\biggr]
\biggr\} , \label{eq:Eunconstrained_expl}
\end{align}
and
\begin{align}
\delta V_\text{pert}(r,\mu_f)=-\frac{4}{3}\,\alpha_s(1/r)\frac{2\mu_f}{\pi}
&\biggl\{
1+\frac{\alpha_s(1/r)}{4\pi}\Bigl(P_1\bigl(\ln\tfrac{1}{r^2\mu_f^2}\bigr)+2\beta_0\Bigr) \nonumber\\
&\hspace*{-2mm}+\biggl(\frac{\alpha_s(1/r)}{4\pi}\biggr)^2\biggl[P_2\bigl(\ln\tfrac{1}{r^2\mu_f^2}\bigr)
+2P_2'\bigl(\ln\tfrac{1}{r^2\mu_f^2}\bigr) +8\beta_0^2\biggr] \nonumber\\
&\hspace*{-2mm}+\biggl(\frac{\alpha_s(1/r)}{4\pi}\biggr)^3\biggl[a_{3{\rm ln}}\ln\alpha_s+a_{3{\rm ln}}\biggl(\frac{1}{2}-\gamma_E-\frac{r\mu_f}{\pi}\biggr)+48\beta_0^3  \nonumber\\
&\hspace*{2.2cm}+\frac{a_{3{\rm ln}}}{2}\ln\bigl(\tfrac{1}{r^2\mu_f^2}\bigr)+ P_3\bigl(\ln\tfrac{1}{r^2\mu_f^2}\bigr)\nonumber\\
&\hspace*{2.2cm}+2P_3'\bigl(\ln\tfrac{1}{r^2\mu_f^2}\bigr) +4P_3''\bigl(\ln\tfrac{1}{r^2\mu_f^2}\bigr) 
\biggr]+{\cal O}(r^2\mu_f^2)\biggr\}. \label{eq:deltaE_expl}
\end{align}
In the latter expression we limited ourselves to the leading term in an expansion in powers of $r\mu_f$ \cite{Beneke:1998rk}.
This is completely sufficient as the terms explicitly accounted for in \Eqref{eq:deltaE_expl} are exactly those canceling the pathological contribution $\sim-\frac{\alpha_s}{r}(r\Lambda_{\rm QCD})$ induced by the Fourier integral~\eqref{eq:Fint} in \Eqref{eq:deltaV}.
Uncontrolled higher-order terms cannot be fully eliminated along these lines anyway.

\subsection{The perturbative coupling $\alpha_s(\mu)$ and its relation to $\Lambda_{\overline{\rm MS}}$} \label{sec:alpha}

Up to now, we did not specify how the strong coupling $\alpha_s(\mu)$ is promoted to an explicit numerical value.
This identification involves the definition of a reference scale, which -- in the ${\overline{\rm MS}}$ scheme -- is denoted by $\Lambda_{\overline{\rm MS}}$.
More specifically, the scale $\Lambda_{\overline{\rm MS}}$ is introduced in terms of specific initial conditions in the integration of \Eqref{eq:beta} \cite{Chetyrkin:1997sg}
(cf. also \cite{Karbstein:2014bsa}),
\begin{equation}
 \frac{\mu}{\Lambda_{\overline{\rm MS}}}=\left(\frac{\beta_0\alpha_s(\mu)}{4\pi}\right)^{\frac{b_1}{2\beta_0}}\exp\biggl\{\frac{2\pi}{\beta_0\alpha_s(\mu)}
 +\frac{1}{\beta_0}\int_0^{\alpha_s(\mu)}\frac{{\rm d}\alpha_s}{\alpha_s}\left(\frac{\beta_0}{\beta(\alpha_s)}+\frac{2\pi}{\alpha_s}-\frac{b_1}{2}\right)\biggr\}. \label{eq:Lambda2-4}
\end{equation}
This scale cannot be determined within perturbation theory, but has to be provided as an external input parameter.
In this study, we aim at determining $\Lambda_{\overline{\rm MS}}^{(n_f=2)}$, i.e., $\Lambda_{\overline{\rm MS}}$ for $n_f=2$ dynamical massless quark flavors, from lattice simulations of the static potential.

Equation~\eqref{eq:Lambda2-4} constitutes an implicit equation for $\alpha_s(\mu)$ as a function of $\mu/\Lambda_{\overline{\rm MS}}$.
Aiming at the best achievable precision, \Eqref{eq:Lambda2-4} can be inverted numerically for $\alpha_s(\mu)$, resorting to two different options, namely by \cite{Karbstein:2014bsa}\footnote{Note 
that the expansion coefficient $b_4$ \cite{Baikov:2016tgj} was not yet 
known in our previous studies~\cite{Jansen:2011vv,Karbstein:2014bsa}.}
\begin{itemize}
 \item[(I)] either plugging the perturbative $\beta$-function~\eqref{eq:betaseries} at the best known accuracy into \Eqref{eq:Lambda2-4} and performing the integration over $\alpha_s$ numerically,
 \item[(II)] or adopting a Taylor expansion of the integrand in \Eqref{eq:Lambda2-4} and performing the integral analytically. To this end only those terms whose coefficients are known explicitly are kept, i.e.,
\begin{multline}
 \int_0^{\alpha_s(\mu)}\frac{{\rm d}\alpha_s}{\alpha_s}\left(\frac{\beta_0}{\beta(\alpha_s)}+\frac{2\pi}{\alpha_s}-\frac{b_1}{2}\right) 
 = \frac{b_2 - b_1^2}{2}\frac{\alpha_s(\mu)}{4\pi} + \frac{b_3 - 2b_1b_2+b_1^3}{4}\biggl(\frac{\alpha_s(\mu)}{4\pi}\biggr)^2 \\ +\frac{b_4-b_2^2-2b_1b_3+3b_1^2b_2-b_1^4}{6}\biggl(\frac{\alpha_s(\mu)}{4\pi}\biggr)^3 +{\cal O}(\alpha_s^4). \label{eq:auch}
\end{multline}
\end{itemize}
These two choices may also serve as a consistency criterion as in the manifestly perturbative regime both options should, of course, yield compatible results.

Besides, \Eqref{eq:Lambda2-4} can be solved approximately for $\alpha_s(\mu)$ \cite{Chetyrkin:1997sg} by first adopting the expansion~\eqref{eq:auch}, then employing a power-series ansatz for $\alpha_s(\mu)$ in powers of $1/\ell$, with $\ell=\ln(\mu^2/\Lambda_{\overline{\rm MS}}^2)$, and finally iteratively determining the expansion coefficients. This results in
\begin{align}
 \alpha_s(\mu)=\frac{4\pi}{\beta_0 l}&\biggl\{1-\frac{b_1}{\beta_0\ell}\ln\ell+\Bigl(\frac{b_1}{\beta_0\ell}\Bigr)^2\biggl(\ln^2\ell-\ln\ell -1+\frac{b_2}{b_1^2}\biggr) \nonumber\\
&-\Bigl(\frac{b_1}{\beta_0\ell}\Bigr)^3\biggl[\ln^3\ell-\frac{5}{2}\ln^2\ell
-\Bigl(2-3\frac{b_2}{b_1^2}\Bigr)\ln\ell +\frac{1}{2}\Bigl(1-\frac{b_3}{b_1^3}\Bigr)\biggr] \nonumber\\
&+\Bigl(\frac{b_1}{\beta_0\ell}\Bigr)^4\biggl[\ln^4\ell-\frac{13}{3}\ln^3\ell-\Bigl(\frac{3}{2}-6\frac{b_2}{b_1^2}\Bigr)\ln^2\ell+\Bigl(4-3\frac{b_2}{b_1^2}-2\frac{b_3}{b_1^3}\Bigr)\ln\ell  \nonumber\\
&\hspace*{4.25cm}+\frac{7}{6}-\frac{b_2}{b_1^2}\Bigl(3-\frac{5}{3}\frac{b_2}{b_1^2}\Bigr)-\frac{1}{6}\frac{b_3}{b_1^3}+\frac{1}{3}\frac{b_4}{b_1^4}\biggr]+{\cal O}\Bigl(\frac{1}{\ell^5}\Bigr)\biggr\} \,. \label{eq:alpha_s}
\end{align}
Terms beyond ${\cal O}(1/\ell^5)$ also include higher, to date unknown coefficients $b_{i\geq 5}$ of the QCD $\beta$-function.

Even though the options (I) and (II) allow for reliable results for $\alpha_s(\mu)$ in a wider range of $\mu$,
\Eqref{eq:alpha_s} is specifically useful when aiming at an analytic expression of the static potential in the limit of $\mu\gg\Lambda_{\overline{\rm MS}}$.
Hence, in our determination of the value of $\Lambda_{\overline{\rm MS}}$ by matching the perturbative static potential to lattice results we will exclusively resort to the options (I) and (II). Contrarily, for the analytic expression of the position space potential we will adopt \Eqref{eq:alpha_s}.


\newpage

\section{\label{SEC547}Determination of $\Lambda_{\overline{\textrm{MS}}}$ for $n_f = 2$ massless quark flavors}

In this section we determine $\Lambda_{\overline{\textrm{MS}}}$ by matching the  perturbative static potential $\tilde{V}_{\rm pert}$ in momentum space, \Eqref{eq:V(p)_known}, to the Fourier transform of our analytic parameterization~\eqref{EQN170} of the lattice static potential $V_{\rm lat}$. 
Recall that the lattice static potential in momentum space allows for reliable insights below a certain momentum, while the perturbative results is applicable above a certain momentum. In turn, the matching procedure has to be done in the momentum regime where the validity of both results overlaps.


\subsection{\label{SEC699}Matching perturbative and lattice QCD results for the static potential}

The analytic parameterization of the lattice potential $V_\textrm{lat}(r)$ in \Eqref{EQN170} can be straightforwardly transformed to momentum space,
\begin{eqnarray}
\tilde{V}_\textrm{lat}(p)  =  \int d^3r \, e^{-i \vec{p}\cdot\vec{r}}\, V_\textrm{lat}(r)  =  -\sigma \frac{8 \pi}{p^4} + \alpha \frac{4 \pi}{p^2} , \label{eq:Vlatt(p)}
\end{eqnarray}
where the physically irrelevant constant $V_0$ has been set to zero.
In this section, we exclusively adopt the ``continuum/combined'' values for the parameters $\alpha$ and $\sigma$ listed in the last row of Table~\ref{TAB766}. 

For the perturbative static potential $\tilde{V}_{\rm pert}(p)$, we adopt the expression given in \Eqref{eq:V(p)_known}, with the implicit equation~\eqref{eq:Lambda2-4} inverted by one of the options (I) or (II) for the strong coupling $\alpha_s(p)$.
In turn, $\tilde{V}_{\rm pert}(p)$ is determined up to a single parameter, namely $\Lambda_{\overline{\textrm{MS}}}^{(n_f=2)}$.

\begin{figure}[htb]
\begin{center}

\begin{picture}(0,0)%
\includegraphics{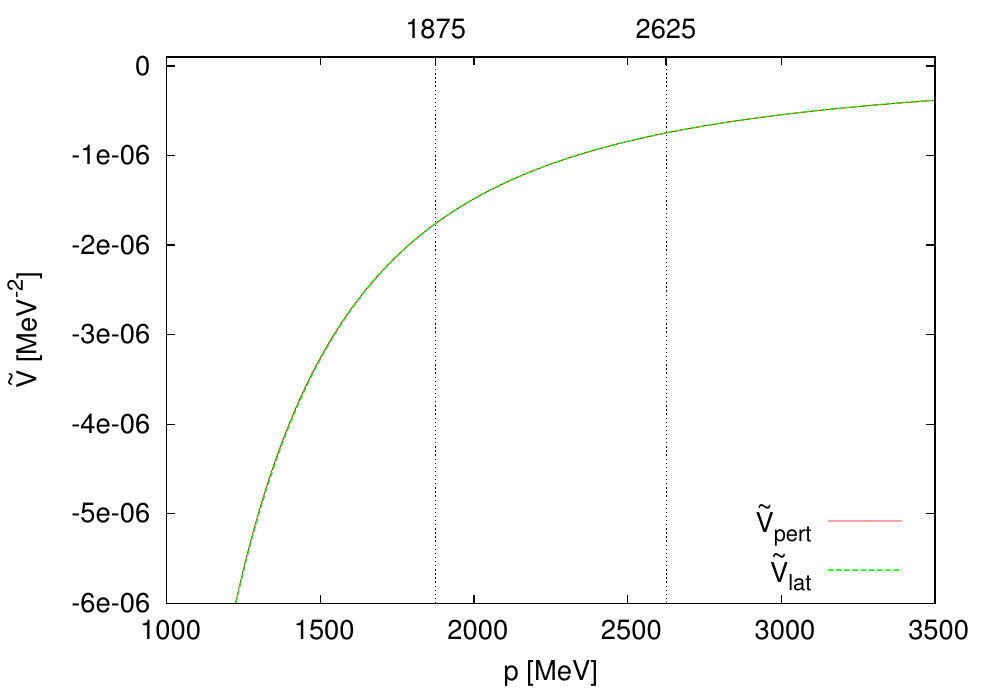}%
\end{picture}%
\setlength{\unitlength}{4144sp}%
\begingroup\makeatletter\ifx\SetFigFont\undefined%
\gdef\SetFigFont#1#2#3#4#5{%
  \reset@font\fontsize{#1}{#2pt}%
  \fontfamily{#3}\fontseries{#4}\fontshape{#5}%
  \selectfont}%
\fi\endgroup%
\begin{picture}(4500,3149)(1,-2310)
\end{picture}%

\caption{\label{FIG005}Matching of $\tilde{V}_{\rm pert}(p)$ in \Eqref{eq:V(p)_known} to $\tilde{V}_{\rm lat}(p)$ in \Eqref{eq:Vlatt(p)} using option~(II), $p_{\min} = 1875 \, \textrm{MeV}$ and $p_{\max} = 2625 \, \textrm{MeV}$. In particular for $p_{\min} \leq p \leq p_{\max}$ the two curves are in perfect agreement.}
\end{center}
\end{figure}

To match $\tilde{V}_\textrm{pert}(p)$ and $\tilde{V}_\textrm{lat}(p)$, we minimize their squared difference,
\begin{equation}
\label{EQN794} \Delta(\Lambda_{\overline{\textrm{MS}}}^{(n_f=2)}) = \int_{p_{\min}}^{p_{\max}} dp \, \Big(\tilde{V}_\textrm{pert}(p) - \tilde{V}_\textrm{lat}(p)\Big)^2 ,
\end{equation}
with respect to $\Lambda_{\overline{\textrm{MS}}}^{(n_f=2)}$ in a given momentum interval $p_{\min} \leq p \leq p_{\max}$.
The lower and upper momenta, $p_{\min}$ and $p_{\max}$, respectively, are chosen such that both $\tilde{V}_\textrm{pert}$ and $\tilde{V}_\textrm{lat}$ exhibit small systematic errors, i.e.\ errors due to the lattice discretization and the truncation of the perturbative series.
The minimization is done numerically using standard integration and root finding techniques.
The small difference in the results obtained by options (I) and (II) is included in the systematic error for the final result (cf.\ the discussion below).
For an exemplarily plot, cf. Figure~\ref{FIG005}.

In our previous determination of $\Lambda_{\overline{\textrm{MS}}}^{(n_f=2)}$ in momentum space \cite{Karbstein:2014bsa} we additionally had to account for a constant offset $\tilde{V}_0$ in the matching of $\tilde{V}_\textrm{pert}(p)$ and $\tilde{V}_\textrm{lat}(p)$. Such a constant was needed in \cite{Karbstein:2014bsa}, because the matching was performed with discrete lattice data points for the static potential in momentum space obtained by means of a discrete Fourier transform from the corresponding lattice data in position space.
Due to lattice discretization errors at small separations $r$, the discrete lattice data points in position space do not exhibit a singularity for $r \to 0$, but form a negative peak, saturating at a finite value. The shape of the peak depends on the lattice spacing $a$ and becomes more pronounced for smaller $a$. Such a peak in position space translates into an $a$-dependent constant in momentum space.
Contrarily, in the present work we parametrize the lattice data points in position space with the function~\eqref{EQN170}, intrinsically excluding lattice data points for small $r$ exhibiting large errors, and thus do not need to account for such a constant shift.
A further argument against the necessity of such a constant shift in the present study is the fact that both $\tilde{V}_\textrm{lat}(p)$ and $\tilde{V}_\textrm{pert}(p)$ exhibit the same asymptotic behavior for $p \rightarrow \infty$. For consistency, we checked this numerically and found $\tilde{V}_0 = 0$ within statistical errors.


\subsection{Variation of input parameters, final result and uncertainty of $\Lambda_{\overline{\textrm{MS}}}$}

To allow a fair comparison of the results, we determine the systematic error of $\Lambda_{\overline{\textrm{MS}}}^{(n_f=2)}$ in the same way as in our previous articles \cite{Jansen:2011vv,Karbstein:2014bsa}. More specifically, we perform the matching procedure outlined in Sec.~\ref{SEC699} $20 \, 000$ times, while varying the input parameters as follows:
\begin{itemize}
\item 50\% of the matching is done with $\tilde{V}_{\rm pert}$ as defined in \Eqref{eq:V(p)_known} at NNLO, 50\% at NNNLO.

\item 50\% of the matching uses option (I) to invert \Eqref{eq:Lambda2-4} for $\alpha_s(p)$, 50\% option (II).

\item For each matching, we randomly choose
\begin{itemize}
\item $p_{\min} \in [1500,2250] \, \textrm{MeV}$,

\item $p_{\max} \in [2250,3000] \, \textrm{MeV}$,
\end{itemize}
with the constraint $p_{\max}-p_{\min} \geq 375 \, \textrm{MeV}$.
\end{itemize}
As this is exactly the procedure used in Sec. 4.2 of our previous work \cite{Karbstein:2014bsa} for the extraction of $\Lambda_{\overline{\textrm{MS}}}^{(n_f=2)}$, though now based on an improved and streamlined determination of $\tilde V_\text{lat}$ (cf. the detailed discussion in Sec.~\ref{SEC462} above), we can resort to findings of \cite{Karbstein:2014bsa}.
This in particular applies for the detailed arguments on the choice of $p_{\min}$ and $p_{\max}$.
For completeness, let us just briefly recall the main points: The lower bound arises from an analyis of the relative importance of the known perturbative orders of $\tilde{V}_\text{pert}$. For $n_f=2$, this motivates the constraint $\alpha_s(p)\lesssim0.3$, translating into $p\gtrsim1500\,{\rm MeV}$.
The upper bound arises from constraints due to lattice discretization effects. The maximum momentum on our finest lattice along an axis, given by $\pi / a \approx 15 \, \textrm{GeV}$, motivates the criterion $p\lesssim \pi/(3a) \approx 5000 \, {\rm MeV}$ for reasonably small lattice discretization errors \cite{Karbstein:2014bsa}. To be on the safe side, we choose the value of $p_{\rm max}$ even somewhat smaller, namely $p_{\rm max}\lesssim3000\,{\rm MeV}$.
In \cite{Karbstein:2014bsa}, we have moreover confirmed the stability of the extracted result for $\Lambda_{\overline{\textrm{MS}}}^{(n_f=2)}$ under variations of $p_{\min}$ and $p_{\max}$, reflecting itself in almost perfect plateaus in Fig.~4 of \cite{Karbstein:2014bsa}.

For each matching we randomly pick one of the jackknife samples $(\alpha,\sigma)$ used in step~(iii) of Sec.~\ref{SEC462} to generate our final ``continuum/combined'' results for $\alpha$ and $\sigma$, listed in the last row of Table~\ref{TAB766}. This ensures that the statistical uncertainties of these parameters and their correlation are properly taken into account.

The impact of finite volume effects on $\Lambda_{\overline{\textrm{MS}}}^{(n_f=2)}$ was studied in \cite{Jansen:2011vv} and found to be negligible in comparison to other errors. Similarly, the effects of non-vanishing light quark masses on $\Lambda_{\overline{\textrm{MS}}}^{(n_f=2)}$ were examined in detail in \cite{Jansen:2011vv} by performing computations with different pion masses in the range $m_\textrm{PS} \approx 325 \, \textrm{MeV} \ldots 517 \, \textrm{MeV}$ at fixed lattice spacing and spacetime volume.
These calculations found $\Lambda_{\overline{\textrm{MS}}}^{(n_f=2)}$ to be stable and constant within tiny statistical errors of $\approx\pm 1 \,\textrm{MeV}$.
Hence, we do not expect the non-vanishing light quark masses in the lattice QCD computation to induce any significant deviations from the limit of massless dynamical quark flavors assumed in the derivation of the perturbative static potential in Sec.~\ref{SEC067}. For these reasons, in the present study we do not account for any potential errors arising from finite volumes and non-vanishing light quark masses on the lattice.

Performing the matching procedure $20 \, 000$ times, we obtain $20 \, 000$ samples for $\Lambda_{\overline{\textrm{MS}}}^{(n_f=2)}$. Their mean value determines our final result, and their standard deviation the corresponding combined statistical and systematic error. This results in
\begin{equation}
\label{EQN863} \Lambda_{\overline{\textrm{MS}}}^{(n_f=2)} = 302(16) \, \textrm{MeV} .
\end{equation}





In comparison to our previous result, $\Lambda_{\overline{\textrm{MS}}}^{(n_f=2)}|_\textrm{\!\!\cite{Karbstein:2014bsa}} = 331(21) \, \textrm{MeV}$, obtained from a momentum space analysis of the static potential in \cite{Karbstein:2014bsa}, our new result is slightly smaller, but still compatible within errors.
More importantly, the error of our new result~\eqref{EQN863} is reduced by roughly 25\% with regard to our previous result, indicating that the procedure presented here allows for a more accurate determination of $\Lambda_{\overline{\textrm{MS}}}$.
We mainly attribute this reduction of the error to the more straightforward and less technically challenging approach pursued for the $\Lambda_{\overline{\textrm{MS}}}$ determination.
Resorting to the continuous parameterization~\eqref{EQN170} of the discrete lattice potential, which has an analytical Fourier momentum space reprentation~\eqref{eq:Vlatt(p)}, technicalities introducing additional systematic errors, such as the cylinder cut employed for the discrete numerical Fourier transform in \cite{Karbstein:2014bsa}, are no longer required.
Another notable difference to our previous determination of $\Lambda_{\overline{\textrm{MS}}}$ \cite{Karbstein:2014bsa} is the inclusion of the coefficient $b_4$ in the perturbative expansion of the QCD $\beta$-function \eqref{eq:betaseries}, which was not yet known in \cite{Karbstein:2014bsa}.
However, in the present analysis we have immediately implemented both changes at the same time, such that their separate impacts on our final error estimate cannot be straightforwardly disentangled.

For completeness, also note the result of our initial determination of $\Lambda_{\overline{\textrm{MS}}}$ from a position space analysis of the quark-antiquark static potential based on the same lattice QCD data \cite{Jansen:2011vv}, yielding $\Lambda_{\overline{\textrm{MS}}}^{(n_f=2)}|_{\!\!\text{\cite{Jansen:2011vv}}} = 315(30) \, \textrm{MeV}$, which is also consistent, but exhibits a larger error. This is related to the observation that our momentum space determinations of $\Lambda_{\overline{\textrm{MS}}}$ are essentially unaffected by variations of the input parameters, while the position space determination exhibits a sizable dependence on the input parameters; cf. also Sec.~4.3.1 of \cite{Jansen:2011vv} and Sec.~4.2.1 of \cite{Karbstein:2014bsa}.

A sizable contribution to the error of our final result~\eqref{EQN863} for $\Lambda_{\overline{\textrm{MS}}}^{(n_f=2)}$ is coming from the different values obtained for $\tilde{V}_{\rm pert}$ at NNLO and NNNLO. Since the perturbative expansion converges quickly in the considered momentum regime (cf. also the detailed study in Sec.~4.2.1 of \cite{Karbstein:2014bsa}), this error can be considered as estimated rather conservatively. In turn, we also provide a result for $\Lambda_{\overline{\textrm{MS}}}^{(n_f=2)}$ obtained by exclusively accounting for $\tilde{V}_{\rm pert}$ at NNNLO, yielding
\begin{equation}
\label{EQN497} \Lambda_{\overline{\textrm{MS}}}^{(n_f=2),\textrm{NNNLO}} = 291(12) \, \textrm{MeV} .
\end{equation}
The mean value of this result, is $11 \, \textrm{MeV}$ smaller than that in \Eqref{EQN863}, based on the combination of both NNLO and NNNLO input. At the same time, the error is reduced by roughly 25\%, which is consistent with the findings of our previous momentum space analysis \cite{Karbstein:2014bsa}, yielding $\Lambda_{\overline{\textrm{MS}}}^{(n_f=2)}|_{\text{\!\!\cite{Karbstein:2014bsa}}} = 331(21) \, \textrm{MeV}$ and $\Lambda_{\overline{\textrm{MS}}}^{(n_f=2),\text{NNNLO}}|_{\text{\!\!\cite{Karbstein:2014bsa}}} = 318(16) \, \textrm{MeV}$.
On the other hand, a more conservative estimate of the error would be to take the difference of the NNLO and the NNNLO results for $\Lambda_{\overline{\textrm{MS}}}^{(n_f=2)}$ as a measure of the uncertainty introduced by the perturbative expansion. Adding this difference to the error quoted in (\ref{EQN497}) in quadrature results in
\begin{equation}
\Lambda_{\overline{\textrm{MS}}}^{(n_f=2),{\rm NNNLO},\Delta} = 291(25) \, \textrm{MeV} .
\end{equation}



\newpage

\section{\label{SEC421}Complete analytic parameterization of the static potential}

The analytic parameterization of the lattice potential $V_{\rm lat}(r)$ derived in \Eqref{EQN170}, and the perturbative static potential $V_{\rm pert}(r)$ as defined in Eqs.~\eqref{eq:deltaV}-\eqref{eq:deltaE_expl}, with the strong coupling given by \Eqref{eq:alpha_s} and $\mu=1/r$, can be combined to provide a complete analytic parameterization of the quark-antiquark static potential $V(r)$ for $n_f=2$ valid up to the string breaking distance.
This parameterization will be useful for various applications, such as heavy-quark phenomenology.


\subsection{Construction of the analytic parameterization of $V(r)$}

The analytic parameterization~\eqref{EQN170} of the lattice potential $V_{\rm lat}(r)$, with parameters $\alpha$ and $\sigma$ fixed to the ``continuum/combined'' results provided in the last row of Table~\ref{TAB766}, accurately describes the quark-antiquark potential for heavy-quark separations $r \gtapprox 3 a$ (cf.\ Table~\ref{TAB766}). For our smallest lattice spacing, the latter condition corresponds to $r \gtapprox 0.12 \, \textrm{fm}$. This expression is valid up to the string breaking distance $r_\textrm{sb} \approx 1 \, \textrm{fm}$.
On the other hand, the perturbative static potential $V_\textrm{pert}(r)$ at NNNLO, Eqs.~\eqref{eq:deltaV}-\eqref{eq:deltaE_expl}, with the strong coupling given by \Eqref{eq:alpha_s}, $\mu=1/r$, $\mu_f \approx 3\ldots7 \Lambda_{\overline{\textrm{MS}}}^{(n_f=2)}$ \cite{Karbstein:2013zxa} and $\Lambda_{\overline{\textrm{MS}}}^{(n_f=2)}$ fixed to the value provided in \Eqref{EQN863}, is expected to exhibit small systematic errors for $r \ltapprox 0.12 \, \textrm{fm}$ \cite{Karbstein:2013zxa}.

To provide an analytic parameterization of $V(r)$ for $r \ltapprox r_\textrm{sb}$, we use these two results and connect them in a smooth way, making use of the so-far undetermined, constant offset $V_0$ between the perturbative and lattice static potentials, accounted for in the definition of $V_\text{lat}$ in \Eqref{EQN170}. This offset originates in the regularization dependent self energy of the static quarks, and naturally differs in perturbation theory and lattice QCD. More specifically, on the lattice the self energy is finite and depends on the lattice spacing and on the smearing of temporal links. More specifically, we define
\begin{eqnarray}
\label{EQN795} V(r)  =  \left\{\begin{array}{cl}
V_\textrm{pert}(r) & \textrm{for } r < r_1 \\
V_{1 2}(r) &\textrm{for } r_1 \leq r \leq r_2 \\
V_\textrm{lat}(r)  & \textrm{for } r_2 < r
\end{array}\right. ,
\end{eqnarray}
where $r_1 , r_2 \approx 0.1 \, \textrm{fm} \ldots 0.2 \, \textrm{fm}$, fulfilling $r_1 < r_2$, and
\begin{eqnarray}
V_{1 2}(r)  =  A + B r + C r^2
\end{eqnarray}
is a second degree polynomial.
The coefficients $A$, $B$, $C$ and the constant shift $V_0$ are chosen such that $V(r)$ is a continuous and smooth ($C^1$ continuous) function, by solving the following system of four simple linear equations,
\begin{eqnarray}
V_\textrm{pert}(r_1) = V_{1 2}(r_1)\,, \quad V'_\textrm{pert}(r_1) = V'_{1 2}(r_1)\,, \quad V_{1 2}(r_2) = V_\textrm{lat}(r_2)\,, \quad V'_{1 2}(r_2) = V'_\textrm{lat}(r_2)\,, \label{eq:Vconds}
\end{eqnarray}
with the primes denoting differentiations with respect to $r$.

\begin{figure}[htb]
\begin{center}

\begin{picture}(0,0)%
\includegraphics{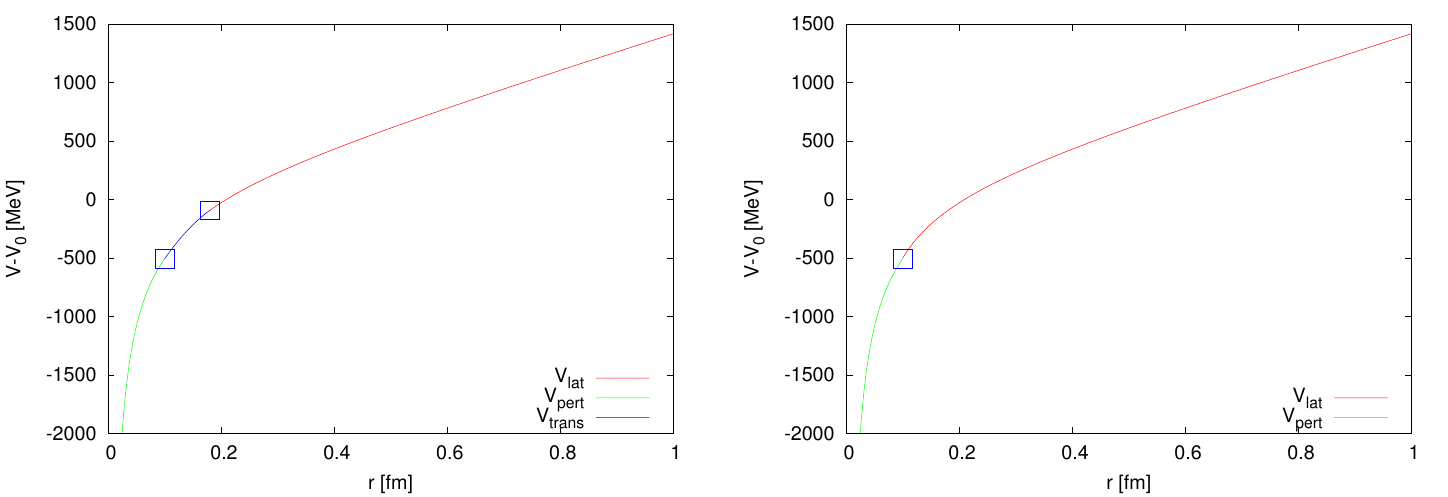}%
\end{picture}%
\setlength{\unitlength}{4144sp}%
\begingroup\makeatletter\ifx\SetFigFont\undefined%
\gdef\SetFigFont#1#2#3#4#5{%
  \reset@font\fontsize{#1}{#2pt}%
  \fontfamily{#3}\fontseries{#4}\fontshape{#5}%
  \selectfont}%
\fi\endgroup%
\begin{picture}(6615,2267)(1,-1428)
\end{picture}%

\caption{\label{FIG006}Analytic parameterization of the static potential for the parameters $\Lambda_{\overline{\textrm{MS}}}^{(n_f=2)} = 302 \, \textrm{MeV}$, $\mu_f = 5 \times 302 \, \textrm{MeV}$, $\alpha = 0.326$ and $\sigma = 7.52 / \textrm{fm}^2$. Left: $V(r)$ constructed along the lines of Eqs.~\eqref{EQN795}-\eqref{eq:Vconds} with $r_1 = 0.10 \, \textrm{fm}$ and $r_2 = 0.18 \, \textrm{fm}$. Right: $V(r)$ obtained by setting $r_1=r_2 = 0.10 \, \textrm{fm}$ in \Eqref{EQN795}.}
\end{center}
\end{figure}

As just mentioned, our motivation to use the second degree polynomial $V_{12}(r)$ in the intermediate region is to provide a smooth analytic parameterization of $V(r)$.
For sufficiently small extents $r_2-r_1$, this choice should accurately describe the static potential in this region.
We believe that the extents $r_2 - r_1 < 0.1 \, \textrm{fm}$ considered here fall into this category. For further evidence, see Figure~\ref{FIG006} (left), constructed via the procedure outlined in Eqs.~\eqref{EQN795}-\eqref{eq:Vconds}.
An alternative, simpler option is to directly connect $V_\textrm{pert}(r)$ and $V_\textrm{lat}(r)$ by choosing $V_0$ appropriately, i.e.\ to set $r_1=r_2$ in \Eqref{EQN795}, thereby omitting the interpolator $V_{12}(r)$.
As an immediate drawback, the so obtained $V(r)$ is not $C^1$ continuous; while $V'_\textrm{pert}(r_1)$ and $V'_\textrm{lat}(r_1)$ are quite similar, they are not identical. 
However, note that the mismatch of perturbation theory and lattice QCD is in fact very mild; cf. Figure~\ref{FIG006} (right) adopting this choice.

\subsection{\label{SEC489}Error analysis, when using the analytic parameterization of $V(r)$}

In Secs.~\ref{SEC068}-\ref{SEC547} possible uncertainties and errors associated with the lattice QCD computation and the perturbative calculation were discussed and quantified in detail. For example we did not only provide individual errors for $\alpha$ and $\sigma$, but also accounted for their correlation.
In this section, we propose the following procedure, allowing for the proper inclusion and propagation of these uncertainties to observables whose determination is based on our complete analytic parameterization of the static potential~\eqref{EQN795} (an explicit example is presented in Sec.~\ref{SEC566}):
\begin{itemize}
\item Let $X$ denote the observable. For example $X$ could be a specific difference of two bottomonium masses, e.g.\ $X \equiv m_{\eta_b(2S)} - m_{\eta_b(1S)}$ (cf. Sec.~\ref{SEC566}).

\item Repeat the calculation of $X$ very often, namely $N\gg100$ times, by randomly sampling the parameters of $V(r)$.
The results are $X_j$, with $j = 1,\ldots,N$.
To this end we choose:
\begin{itemize}
\item $\alpha$ and $\sigma$ according to the following 2-dimensional Gaussian probability distribution, characterized by the parameters $\overline{\alpha}$, $\Delta \alpha$, $\overline{\sigma}$, $\Delta \sigma$ and $\textrm{corr}(\alpha,\sigma)$ of the ``continuum/combined'' results listed in the last row of Table~\ref{TAB766},
\begin{align}
\nonumber p(\alpha,\sigma) &  = \frac{1}{2 \pi \Delta \alpha \Delta \sigma \sqrt{1 - \textrm{corr}(\alpha,\sigma)^2}} \exp\Biggl\{-\frac{1}{2 (1 - \textrm{corr}(\alpha,\sigma)^2)} \\
 & \hspace{0.675cm}
\biggl(\!\begin{array}{c} (\alpha - \overline{\alpha}) / \Delta \alpha \\ (\sigma - \overline{\sigma}) / \Delta \sigma \end{array}\!\biggr)^T
\biggl(\!\begin{array}{cc} 1 & -\textrm{corr}(\alpha,\sigma) \\ -\textrm{corr}(\alpha,\sigma) & 1 \end{array}\!\biggr)
\biggl(\!\begin{array}{c} (\alpha - \overline{\alpha}) / \Delta \alpha \\ (\sigma - \overline{\sigma}) / \Delta \sigma \end{array}\!\biggr)
\Biggr\} .
\end{align}

\item $\Lambda_{\overline{\textrm{MS}}}^{(n_f=2)}$ according to a Gaussian probability distribution parameterized by our result in \Eqref{EQN863}. For completeness, note that in principle, $\Lambda_{\overline{\textrm{MS}}}^{(n_f=2)}$, $\alpha$ and $\sigma$ are also correlated. As this correlation is rather small, we neglect it in the error analysis.

\item $\mu_f \in [3 \, \Lambda_{\overline{\textrm{MS}}}^{(n_f=2)} , 7 \, \Lambda_{\overline{\textrm{MS}}}^{(n_f=2)}]$ uniformly.

\item $r_1 \in [0.08 \, \textrm{fm} , 0.12 \, \textrm{fm}]$, $r_2 \in [0.16 \, \textrm{fm} , 0.20 \, \textrm{fm}]$, uniformly.
\end{itemize}

\item The mean value of the obtained results $X_j$ serves as our estimate for $X$. Its error is defined via the standard deviation. 
More specifically,
\begin{eqnarray}
\overline{X}  =  \frac{1}{N} \sum_{j=1}^N X_j \, , \quad \Delta X  =  \bigg(\frac{1}{N} \sum_{j=1}^N \Big(X_j - \overline{X}\Big)^2\bigg)^{1/2} .
\end{eqnarray}

\item Check that both $\overline{X}$ and $\Delta X$ are essentially independent of $N$. If not, increase $N$.
\end{itemize}


\newpage

\section{\label{SEC566}The bottomonium spectrum in the Born-Oppenheimer approximation}

In the following, we use the parameterization of the static potential~\eqref{EQN795} to compute the bottomonium spectrum. To this end, we employ the Born-Oppenheimer approximation \cite{Born:1927}, which consists of two steps. Its first step amounts to the computation of the static potential, assuming the light quarks and gluons as dynamical degrees of freedom and the $b$ quark and its antiquark $\bar{b}$ as static.
In the second step, this constraint is relaxed, and the Schr\"odinger equation for the relative coordinate of the $\bar{b} b$ pair is solved with the potential computed in the first step, assuming a finite $b$ quark mass $m_b$.

Since the static potential manifestly neglects $1/m_b$ corrections, encoding e.g. spin effects of the heavy quarks, this approach certainly does not allow us to obtain very accurate results for the bottomonium spectrum.
We rather intend at performing an exemplarily calculation, utilizing the parameterization~\eqref{EQN795} of the static potential and including a full error propagation along the lines of Sec.~\ref{SEC489}.
Moreover, it can be considered as a preparatory step for a more refined computation, accounting for such $1/m_b$ corrections, to be determined within potential non-relativistic QCD and lattice QCD (cf.\ e.g.\ \cite{Brambilla:2000gk,Pineda:2000sz,Koma:2006si,Koma:2006fw,Koma:2007jq}). Also note, that the Born-Oppenheimer approach without $1 / m_b$ corrections has recently been used for the study of heavy tetraquarks (cf.\ e.g.\ \cite{Bicudo:2012qt,Brown:2012tm,Bicudo:2015vta,Bicudo:2015kna,Peters:2016wjm,Bicudo:2016ooe,Bicudo:2017szl}), where no experimental data is available.
A comparison of the theoretical predictions for the bottomonium masses with corresponding experimental results, might provide us with an estimate of the systematic error associated with this approach.

To this end, we solve the Schr\"odinger equation
\begin{eqnarray}
\bigg(-\frac{1}{2 \mu} \triangle + V(r)\bigg) \psi(\vec{r})  =  E \psi(\vec{r}) , \label{eq:S1}
\end{eqnarray}
where $\vec{r}$ is the relative coordinate of the $b \bar{b}$ pair, $V(r)$ is the parameterization~\eqref{EQN795} of the static potential and $\mu = m_b / 2$ the reduced mass of the $b$ quark.
For its explicit value, we employ either $m_b = m_{b,\overline{\textrm{MS}}} = 4.18 \, \textrm{GeV}$ determined in the $\overline{\textrm{MS}}$ scheme \cite{Patrignani:2016xqp}, or $m_b = m_{b,\textrm{qm}} = 4.977 \, \textrm{GeV}$ from quark models \cite{Godfrey:1985xj}.
Since the potential is radially symmetric, \Eqref{eq:S1} can be separated in a radial and an angular equation, with the latter being trivial to solve. The radial equation reads
\begin{eqnarray}
\bigg(-\frac{1}{2 \mu} \frac{d^2}{dr^2} + \frac{l (l+1)}{2 \mu r^2} + V(r)\bigg) u_{n,l}(r)  =  E_{n,l} u_{n,l}(r) ,
\end{eqnarray}
where we made use of the ansatz $\psi(\vec{r}) = (u_{n,l}(r) / r) Y_{l,m}(\vartheta,\varphi)$. The solutions $u_{n,l}(r)$ are labeled by the principal quantum number $n$ and the azimuthal quantum number $l$, corresponding to the orbital angular momentum of the $\bar{b} b$ pair. This equation can be solved numerically using standard techniques. Here we use a 4th order Runge-Kutta shooting method combined with Newton's method for root finding.

The resulting energy eigenvalues $E_{n,l}$ can be related to bottomonium masses $M_{n,l}$ via
\begin{eqnarray}
M_{n,l}  =  E_{n,l} - E_{1,0} + m_{\eta_b(1S)} ,
\end{eqnarray}
where $m_{\eta_b(1S)} = 9399 \, \textrm{MeV}$ is fixed by experimental input \cite{Patrignani:2016xqp}. Clearly, experimental input is needed to calibrate the unknown constant shift between the energy eigenvalues $E_{n,l}$ and the bottomonium masses $M_{n,l}$, which has its origin in the self energy of the static quarks and the lattice cutoff.
In other words, using the Born-Oppenheimer approach one can only predict mass differences, but not absolute masses of bottomonium states.
Moreover, the results are independent of the heavy quark spins, because the potential $V(r)$ has been computed in the static limit.
For example, for orbital angular momentum $l = 0$ there are degenerate $J = 0$ and $J = 1$ states, and for $l = 1$ there are degenerate $J = l-1$, $J = l$ and $J = l+1$ states, with $J$ denoting the total angular momentum of the bottomonium system.

In the following we present results for three different cases:
\begin{itemize}
\item[(A)] $m_b = m_{b,\overline{\textrm{MS}}}$, $V(r)=V(r)\big|_{\text{Eq.~}\eqref{EQN795}}$.

\item[(B)] $m_b = m_{b,\textrm{qm}}$, $V(r)=V(r)\big|_{\text{Eq.~}\eqref{EQN795}}$.

\item[(C)] $m_b = m_{b,\textrm{qm}}$,
\begin{eqnarray}
\label{EQN638} V(r)  =  \left\{\begin{array}{cl}
V(r)\big|_{\text{Eq.~}\eqref{EQN795}} & \textrm{for}\ r < r_{\textrm{sb}} \\
V(r_{\textrm{sb}}) = \textrm{const.} &\textrm{for}\ r \geq r_{\textrm{sb}} 
\end{array}\right.,
\end{eqnarray}
where $r_{\textrm{sb}} = 1.13 \, \textrm{fm}$ is the string breaking distance determined with lattice QCD in \cite{Bali:2005fu}.
\end{itemize}
Cases (A) and (B) allow us to infer, how strong bottomonium mass differences depend on the value of $m_b$.\footnote{Another option would be to tune $m_b$ such that for a specific bottomonium mass difference, e.g.\ between the $\eta_b(1S)$ and $\eta_b(2S)$ states, the theoretical result agrees with experiment.}
For cases (B) and (C) we keep the mass of the $b$ quark fixed, and only alter the long-distance behavior of the static potential. This allows us to check whether string breaking effects have a sizable effect on the bottomonium spectrum.
Of course, with the potential~\eqref{EQN638} and the approach adopted here, we can only determine bottomonium states below the $b \bar{b}$ threshold.
States above are unstable resonances. In principle, the determination of such states is possible along the lines, but requires techniques from scattering theory (cf.\ e.g.\ \cite{Bicudo:2017szl}).

Our results for the mass differences $\Delta E_{n,l} = E_{n,l} - E_{1,0}$ with $l\in\{0,1\}$ are collected in Table~\ref{TAB763}. The statistical analysis has been performed with the method detailed in Sec.~\ref{SEC489}, employing $N = 3000$ samples for each bottomonium mass.

\begin{table}[htb]
\begin{center}
\begin{tabular}{|c|cccc|}
\hline
 & & & &\vspace{-0.40cm} \\
 &  & (A) & (B) & (C) \\
 & & & &\vspace{-0.40cm} \\
\hline
 & & & &\vspace{-0.40cm} \\
 \multirow{3}{*}{$l=0$}& $\Delta E_{20}$ & $0.61(2)$ & $0.59(2)$ & $0.59(2)$\\
 & & & &\vspace{-0.40cm} \\
                       & $\Delta E_{30}$ & $1.05(3)$ & $1.02(3)$ & $1.01(2)$\\
 & & & &\vspace{-0.40cm} \\
                       & $\Delta E_{40}$ & $1.43(4)$ & $1.38(4)$ & --\\
 & & & &\vspace{-0.40cm} \\
\hline
 & & & &\vspace{-0.40cm} \\
 \multirow{3}{*}{$l=1$}& $\Delta E_{11}$ & $0.43(1)$ & $0.42(1)$ & $0.43(1)$\\
 & & & &\vspace{-0.40cm} \\
                        & $\Delta E_{21}$ & $0.88(3)$ & $0.87(3)$ & $0.87(3)$\\
 & & & &\vspace{-0.40cm} \\
                       & $\Delta E_{31}$ & $1.26(4)$ & $1.24(4)$ & --\vspace{-0.40cm}\\
 & & & &\\
\hline
\end{tabular}
\caption{\label{TAB763}Mass differences $\Delta E_{n,l} = E_{n,l} - E_{1,0}$ in units of $\textrm{GeV}$ for the three cases (A), (B) and (C) discussed in the main text.}
\end{center}
\end{table}

\begin{figure}[h!]
\begin{center}

\begin{picture}(0,0)%
\includegraphics{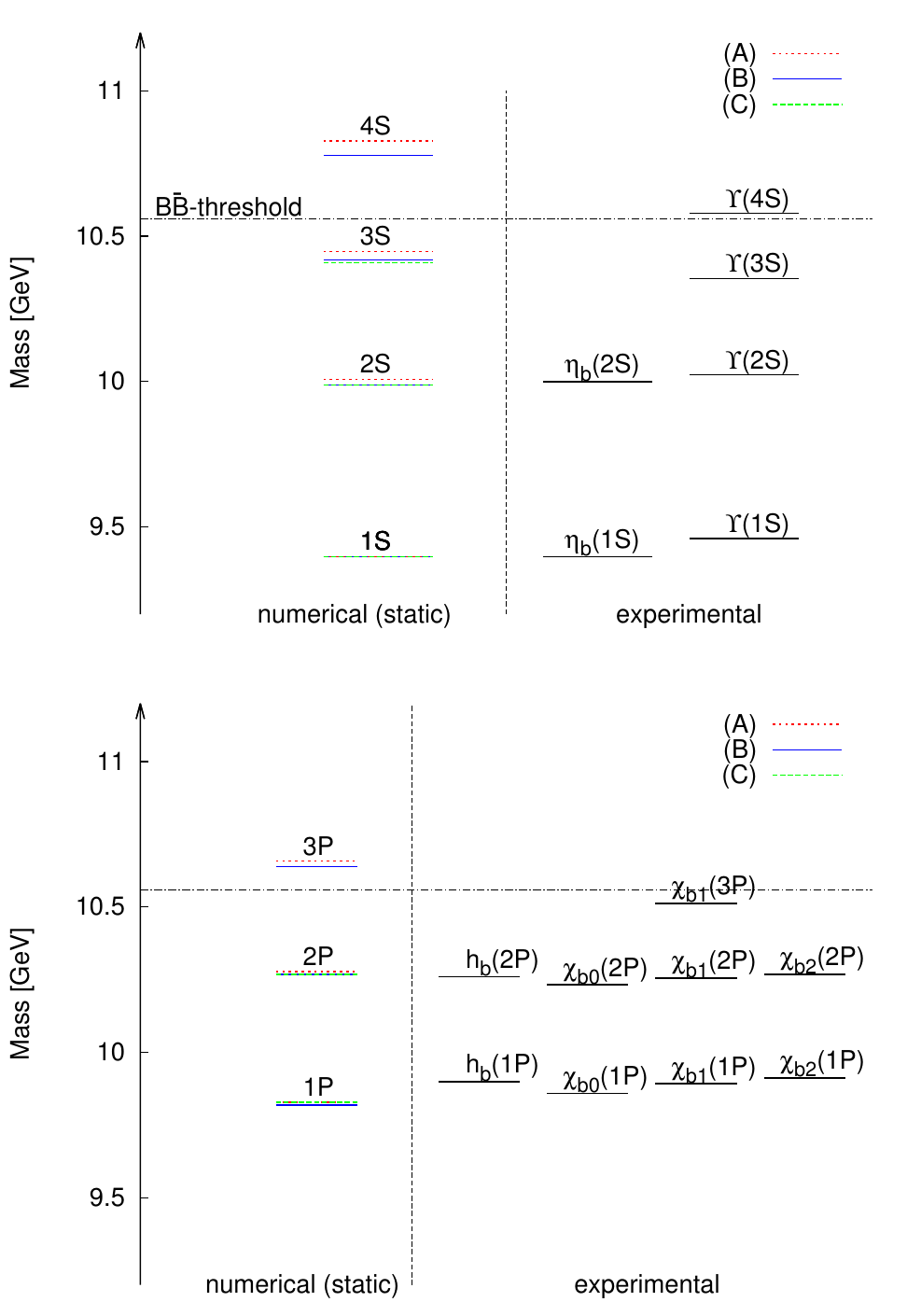}%
\end{picture}%
\setlength{\unitlength}{4144sp}%
\begingroup\makeatletter\ifx\SetFigFont\undefined%
\gdef\SetFigFont#1#2#3#4#5{%
  \reset@font\fontsize{#1}{#2pt}%
  \fontfamily{#3}\fontseries{#4}\fontshape{#5}%
  \selectfont}%
\fi\endgroup%
\begin{picture}(4500,6435)(1,-5596)
\end{picture}%

\caption{\label{FIG641}Masses of bottomonium states in units of $\textrm{GeV}$: graphical representation of the data assembled in Table~\ref{TAB944}. Our theoretical predictions for the three cases (A), (B) and (C) discussed in the main text are confronted with experimental data \cite{Patrignani:2016xqp}. In our numerical computations, the mass of the $1S$ state is fixed to the state $\eta_b(1S)$ observed in experiment.}
\end{center}
\end{figure}

The resulting bottomonium masses are confronted with experimental data in Table~\ref{TAB944} and Figure~\ref{FIG641}, where the common notation $S$ for $L=0$ and $P$ for $L=1$ orbital angular momentum is used. As discussed above, our computations do not account for the heavy quark spins, such that, e.g.\ the bottomonium states $\eta_b(1S)$ and $\Upsilon(1S)$ with quantum numbers $J^{P} = 0^-$ and $J^P = 1^-$, respectively, are mass degenerate. Here, the upper label $P$ refers to the parity of the state. In turn, the experimentally measured mass difference of the order of $50 \, \textrm{MeV}$ for these two states can serve as an estimate of the systematic error associated with our results.

\begin{table}[htb]
\begin{center}
\begin{tabular}{|cc|cccc|}
\hline
 & & & & &\vspace{-0.40cm} \\
  & $n^{2S+1}L_J$ & PDG & (A) & (B) & (C)\\
 & & & & &\vspace{-0.40cm} \\
\hline
\hline
 & & & & &\vspace{-0.40cm} \\
                      $\eta_b(1S)$ & $1^1S_0$ & $\phantom{0}9.399(3)\phantom{00}$ & \multirow{2}{*}{$\phantom{0}9.399(3)^\ast$} & \multirow{2}{*}{$\phantom{0}9.399(3)^\ast$} & \multirow{2}{*}{$\phantom{0}9.399(3)^\ast$} \\
 & & & & &\vspace{-0.40cm} \\
                      $\Upsilon(1S)$ & $1^3S_1$ & $\phantom{0}9.4603(3)\phantom{0}$ &  &  & \\
 & & & & &\vspace{-0.40cm} \\
\hline
\hline
 & & & & &\vspace{-0.40cm} \\
                      $h_b(1P)$ & $1^1P_1$ & $\phantom{0}9.8993(8)\phantom{0}$ & \multirow{6}{*}{$\phantom{0}9.83(1)\phantom{0^\ast}$} & \multirow{6}{*}{$\phantom{0}9.82(1)\phantom{0^\ast}$} & \multirow{6}{*}{$\phantom{0}9.83(1)\phantom{0^\ast}$}\\
 & & & & &\vspace{-0.40cm} \\
 & & & & &\vspace{-0.40cm} \\
                      $\chi_{b0}(1P)$ & $1^3P_0$ & $\phantom{0}9.8594(5)\phantom{0}$ &  &  & \\
 & & & & &\vspace{-0.40cm} \\
 & & & & &\vspace{-0.40cm} \\
                      $\chi_{b1}(1P)$ & $1^3P_1$ & $\phantom{0}9.8928(4)\phantom{0}$ &  &  & \\
 & & & & &\vspace{-0.40cm} \\
 & & & & &\vspace{-0.40cm} \\
                      $\chi_{b2}(1P)$ & $1^3P_2$ & $\phantom{0}9.9122(4)\phantom{0}$ &  &  & \\
 & & & & &\vspace{-0.40cm} \\
\hline
\hline
 & & & & &\vspace{-0.40cm} \\
                      $\eta_b(2S)$ & $2^1S_0$ & $\phantom{0}9.999(4)\phantom{00}$ & \multirow{2}{*}{$10.01(2)\phantom{0^\ast}$} & \multirow{2}{*}{\phantom{0}$9.99(2)\phantom{0^\ast}$} & \multirow{2}{*}{\phantom{0}$9.99(2)\phantom{0^\ast}$}\\
 & & & & &\vspace{-0.40cm} \\
                      $\Upsilon(2S)$ & $2^3S_1$ & $10.0233(3)\phantom{0}$ &  &  & \\
 & & & & &\vspace{-0.40cm} \\
\hline
\hline
 & & & & &\vspace{-0.40cm} \\
                      $h_b(2P)$ & $2^1P_1$ & $10.2598(12)$ & \multirow{6}{*}{$10.28(3)\phantom{0^\ast}$} & \multirow{6}{*}{$10.27(3)\phantom{0^\ast}$} & \multirow{6}{*}{$10.27(3)\phantom{0^\ast}$}\\
 & & & & &\vspace{-0.40cm} \\                  
 & & & & &\vspace{-0.40cm} \\
                      $\chi_{b0}(2P)$ & $2^3P_0$ & $10.2325(6)\phantom{0}$ &  &  & \\
 & & & & &\vspace{-0.40cm} \\
 & & & & &\vspace{-0.40cm} \\
                      $\chi_{b1}(2P)$ & $2^3P_1$ & $10.2555(6)\phantom{0}$ &  &  & \\
 & & & & &\vspace{-0.40cm} \\
 & & & & &\vspace{-0.40cm} \\
                      $\chi_{b2}(2P)$ & $2^3P_2$ & $10.2687(6)\phantom{0}$ &  &  & \\
 & & & & &\vspace{-0.40cm} \\
\hline
\hline
 & & & & &\vspace{-0.40cm} \\
                      $\Upsilon(3S)$ & $3^3S_1$ & $10.3552(5)\phantom{0}$ & $10.45(3)\phantom{0^\ast}$ & $10.42(3)\phantom{0^\ast}$ & $10.41(2)\phantom{0^\ast}$\\
 & & & & &\vspace{-0.40cm} \\
 \hline
 \hline
 & & & & &\vspace{-0.40cm} \\
                      $\chi_{b1}(3P)$ & $3^3P_1$ & $10.5121(23)$ & $10.66(4)\phantom{0^\ast}$ & $10.64(4)\phantom{0^\ast}$ & --\\
 & & & & &\vspace{-0.40cm}\\
\hline
\hline
& & & & &\vspace{-0.40cm} \\
                      $\Upsilon(4S)$ & $4^3S_1$ & $10.5794(12)$ & $10.83(4)\phantom{0^\ast}$ & $10.78(4)\phantom{0^\ast}$ & --\vspace{-0.40cm}\\
 & & & & & \\
\hline
\end{tabular}
\caption{\label{TAB944}Masses of bottomonium states in units of $\textrm{GeV}$: experimental data listed by the Particle Data Group (PDG) \cite{Patrignani:2016xqp} are compared with theoretical predictions for the three cases (A), (B) and (C) discussed in the main text. The values marked with ${}^\ast$ serve as input and are no predictions.}
\end{center}
\end{table}

For the low-lying states $1S$, $2S$, $1P$ and $2P$ our theoretical predictions are in good agreement with experiment within the expected systematic error of the order of $50 \, \textrm{MeV}$. Higher states, in particular those above the $b \bar{b}$ threshold, should be treated with caution.
Noteworthily, all masses below threshold, including those in its vicinity, as e.g. the $3S$ state, are essentially identical for cases (B) and (C), which implies that they are not affected by the flattening of the potential for heavy quark separations above the string breaking distance.

Let us also note, that the static bottomonium results of \cite{Laschka:2011zr} are compatible with ours.
Reference \cite{Laschka:2011zr} does not account for string breaking effects, but models contributions inversely proportional to the heavy quark masses in the potential, encoding spin effects of the heavy quarks and hyperfine splitting. A possible future direction could be to compute $1/m_b$ and $1/m_b^2$ corrections of the potential using lattice QCD, as proposed and pioneered for $n_f = 0$ in \cite{Brambilla:2000gk,Pineda:2000sz,Koma:2006si,Koma:2006fw,Koma:2007jq}.


\newpage

\section{Conclusions and outlook}\label{sec:concl}

In this article, we have determined the parameter $\Lambda_{\overline{\textrm{MS}}}$ for QCD with $n_f=2$ dynamical quark flavors by fitting the perturbative result for the static potential to lattice data in momentum space.
Building on insights from our previous determinations \cite{Jansen:2011vv,Karbstein:2014bsa}, we have substantially improved and streamlined our strategy to extract the value of $\Lambda_{\overline{\textrm{MS}}}$, resulting in
\begin{equation}
 \Lambda_{\overline{\textrm{MS}}}^{(n_f=2)} = 302(16) \, \textrm{MeV} .
\end{equation}
One of the main improvements devised in the present work is the use of an analytic parameterization of the discrete simulation data of the lattice static potential in position space. 
This renders complicated and time-consuming numerical techniques employed in our previous work \cite{Karbstein:2014bsa} superfluous, such as the discrete Fourier transform combined with a cylinder cut, which possibly introduces large systematic errors.
Besides, it immediately provides an analytical expression for the static quark-antiquark potential in the manifestly non-perturbative regime.

In a second step, we have used $\Lambda_{\overline{\textrm{MS}}}$ as input parameter in the perturbative static potential. Utilizing an approximate analytical expression for the strong coupling $\alpha_s(\mu)$ in terms of the dimensionless ratio $\mu/\Lambda_{\overline{\textrm{MS}}}$ valid for large values of $\mu$, upon identification of $\mu = 1/r$, the perturbative static potential in position space becomes an analytic function of the quark-antiquark separation $r$.
This function accurately describes the small distance behavior of the static potential.  
Connecting it with the analytic parameterization of the lattice potential by means of an adequately chosen interpolating function, we have constructed a complete analytic parameterization of the static quark-antiquark potential in position space, valid up to the string breaking distance.
If desired, the effect of string breaking can also be phenomenologically accounted for by letting the potential become constant beyond the string breaking distance or by using first principles lattice QCD input, e.g.\ from \cite{Bali:2005fu}.
This all-distance potential encoding both perturbative and manifestly non-perturbative information has ample phenomenological applications.

As an immediate phenomenological application and example, we have used this potential to determine the bottomonium spectrum in the static limit, based on the Born-Oppenheimer approximation. 
Fixing the lowest bound state with data provided by the Particle Data Group \cite{Patrignani:2016xqp}, all bound states below the $b\bar{b}$ threshold are in reasonable agreement with experiment.
Note that small deviations are not surprising as in particular spin effects have been completely ignored in our current analysis.

Let us finally emphasize, that the strategy devised in the present work can be readily adopted to the determination of $\Lambda_{\overline{\textrm{MS}}}$, and for the construction of analytic all-distance potentials from other lattice configurations with, e.g., $n_f = 0$, $n_f = 2+1$ and $n_f = 2+1+1$ dynamical quark flavors.


\newpage

\section*{Acknowledgments}

We acknowledge useful conversations with Nora Brambilla, Antje Peters and Antonio Vairo. Besides, we thank K\"onig M.\ Dillig for important remarks concerning the color sector of QCD.

This work was supported in part by the Helmholtz International Center for FAIR within the framework of the LOEWE program launched by the State of Hesse.

Calculations on the LOEWE-CSC and FUCHS-CSC high-performance computers of Frankfurt University were conducted for this research. We would like to thank HPC-Hessen, funded by the State Ministry of Higher Education, Research and the Arts, for programming advice.



\end{document}